 \documentclass[12pt, draftclsnofoot, onecolumn]{IEEEtran}

\IEEEoverridecommandlockouts
\usepackage{cite}
\usepackage{amsmath,amssymb,amsfonts}
\usepackage{algorithmic}
\usepackage{graphicx}
\usepackage{textcomp}
\usepackage{xcolor}
\usepackage[utf8]{inputenc}
\usepackage{amsmath}
\usepackage[english]{babel}
\usepackage{amsthm}
\usepackage{mathtools}
\usepackage{multirow}
\usepackage{mathtools}
\usepackage[ruled,norelsize]{algorithm2e}
\usepackage{bm} 
\usepackage{tikz}
\usepackage{subcaption}

\theoremstyle{definition}

\DeclarePairedDelimiter\floor{\lfloor}{\rfloor}

\def\BibTeX{{\rm B\kern-.05em{\sc i\kern-.025em b}\kern-.08em
    T\kern-.1667em\lower.7ex\hbox{E}\kern-.125emX}}
 \newtheorem*{remark}{Remark}

\makeatletter
\newcommand{\removelatexerror}{\let\@latex@error\@gobble}

\begin{document}

\title{{ Modulation Based On A Simple MDS Code: Achieving Better Error Performance Than Index Modulation and Related Schemes}}

\author{Ferhat Yarkin~\IEEEmembership{Student~Member,~IEEE} and Justin P.~Coon~\IEEEmembership{Senior~Member,~IEEE}
\thanks{F. Yarkin and J. P. Coon are with the Department of Engineering Science, University of Oxford, Parks Road, Oxford, OX1 3PJ, U.K. E-mail: \{ferhat.yarkin and justin.coon\}@eng.ox.ac.uk}
\thanks{The authors acknowledge the support of the Bristol Innovation \& Research Laboratory of Toshiba Research Europe Ltd.}
}

\maketitle

\begin{abstract}
 In this paper, we propose two novel modulation concepts based on a simple maximum distance separable (MDS) code { and show that these concepts can achieve better error performance than index modulation (IM) and related schemes.} In the first concept, we use amplitude and phase levels to form a simple MDS code, whereas, in the second one, in-phase and quadrature components of codeword elements are used to construct the MDS code. We depict practical schemes for using the proposed concepts with orthogonal frequency division multiplexing (OFDM). We analyze the performance in terms of the minimum Euclidean distance and bit error rate. We also show that the proposed techniques exhibit desirable properties such as efficient low-complexity detection, very simple bits-to-symbols, and symbols-to-bits mappings, and a better error performance when compared to the OFDM-IM and related schemes. More importantly, contrary to the vast majority of IM studies that focus on showing the superiority of the IM techniques against conventional modulation techniques, we show that modulation concepts based on a well-known MDS code can achieve better error performance than the IM and related schemes while exhibiting a structure as simple as these schemes.
\end{abstract}


\section{Introduction}

Desirable features of a constellation can be listed as high signal-to-noise ratio (SNR) efficiency, low-complexity decoding, simple bits-to-symbols, and symbols-to-bits mappings, compatibility with existing coding and modulation techniques \cite{Forney89}. { In this context,  index modulation (IM) which is a subclass of permutation modulation (PM) in \cite{Slepian65} exhibits desirable properties since its combinatorial structure enables a higher SNR efficiency against conventional techniques, low-complexity implementation, and compatibility with the existing coding and modulation techniques \cite{Ishikawa2018,Mesleh2008,Basar2013,Mao2019,Choi2017,Zhang2021}.} In \cite{Mesleh2008}, a space domain IM concept called spatial modulation has been proposed and shown to perform better than conventional modulation concepts such as QAM and PSK in terms of error performance. The authors of \cite{Basar2013} have introduced a frequency domain application of IM called orthogonal frequency division multiplexing with IM (OFDM-IM) and show that OFDM-IM exhibits a low  implementation complexity as well as a high SNR efficiency when compared to conventional OFDM. The potential of IM has also been documented for other domains such as time and code domains \cite{Mao2019}. Moreover, several studies showed that IM works well with channel coding techniques \cite{Choi2017,Zhang2021}.

For a multi-dimensional constellation operating in fading channels, it is possible to achieve high diversity orders by choosing constellation points suitably \cite{Girand1996,Inacio2020} or by using well-known techniques such as signal space diversity \cite{Boutros98}. However, these techniques trade error performance for system complexity. Thus, they become impractical when we try to achieve high spectral efficiencies (SEs). In this regard, although IM itself cannot attain diversity orders comparable to those of the high diversity schemes, its properties such as low-complexity implementation, higher SNR efficiency than the conventional techniques, and compatibility with the coding techniques introduce a trade-off between system performance and simplicity. Moreover, those properties suffice to make it a promising candidate for next-generation communication networks \cite{Tusha2021,Basar2017ac}. 

On the other hand, conventional IM forms its constellation by activating a certain amount of codeword elements and embedding information into combinations of the activated elements, or in other words index symbols, and conventional modulation symbols. Moreover, the minimum Hamming distance between index symbols { of conventional IM in \cite{Basar2013}} is two, whereas such a distance is limited to one for the modulation symbols. Due to index symbols, one can embed less information into conventional modulation symbols and increase the minimum Euclidean distance (MED) for these symbols. Although this provides an efficient constellation structure for IM, the number of index symbols is limited in IM and nulling some codeword elements limits the spectral and SNR efficiency. { One can also achieve a higher minimum Hamming distance between index symbols, thus a better error performance for IM by carefully choosing the indices of active codeword elements as in \cite{Ishikawa2019}. However, in this case, the number of index symbols reduces and the SE is sacrificed.} To overcome these limitations, one can use distinguishable constellations instead of nulling codeword elements and show that the resulting constellation can achieve a higher spectral and SNR efficiency than the original IM \cite{Mao2017, Wen2017,Yarkin2020set,Yarkin2020}.\footnote{Apart from using distinguishable constellations, one can also use the in-phase and quadrature components, energies, and layers of codeword elements in a combinatorial fashion to improve the performance of IM as shown in \cite{Fan2015,Yarkin2020comp,Yarkin2021,li2019}.} In \cite{Mao2017}, dual-mode OFDM with IM (DM-OFDM-IM) was proposed: this method uses the combinations of two distinguishable\footnote{We use the terms ``distinguishable'' and ``disjoint'' interchangeably in the rest of the paper.} constellations to encode information, whereas the authors of \cite{Wen2017} proposed an OFDM scheme called multi-mode OFDM-IM (MM-OFDM-IM) that utilizes the permutations of the distinguishable constellations instead, thus provides a higher number of index symbols than DM-OFDM-IM and OFDM-IM.\footnote{{It is important to note that just like OFDM-IM, the DM-OFDM-IM and MM-OFDM-IM concepts are the PM concepts since they utilize the combinatorial tools like combinations and permutations in the same way  as the classic PM in \cite{Slepian65} to form their codewords \cite{Ishikawa2018}.}} They also showed MM-OFDM-IM is capable of achieving a considerably better error performance than OFDM-IM and DM-OFDM-IM while preserving a low implementation complexity. Moreover, in \cite{Yarkin2020set}, we uncovered a novel application of set partitioning to embed information into the indices of disjoint constellations and showed that OFDM with set partition modulation (OFDM-SPM) is capable of achieving a higher SE and better error performance than MM-OFDM-IM. In \cite{Yarkin2020}, we propose an MDS coding based modulation concept that uses distinguishable constellations to specify $N$-tuples of an MDS code unlike OFDM-SPM, MM-OFDM-IM, and DM-OFDM-IM that exploit them to construct set partitions, permutations and combinations of codeword elements, respectively. In this way, we achieve the highest number of symbols that have the same minimum Hamming distance as those of index symbols. We further show that the proposed scheme outperforms the OFDM-IM benchmarks in terms of SE and error performance.

Motivated by the desirable properties of the MDS approach in \cite{Yarkin2020},  we generalize this approach in this paper. { Like \cite{Yarkin2020}, we use the MDS approach to obtain the highest number of codewords whose minimum Hamming distance is two just like the index symbols of the IM benchmarks. Moreover, we aim to show that one can use a simple MDS code and obtain practical, flexible, and low-complexity modulation concepts that are capable of outperforming the IM benchmarks.} Unlike \cite{Yarkin2020} that uses disjoint PSK and QAM constellations to form a simple MDS code, we take a more general approach that we exploit amplitude and phase levels as well as in-phase and quadrature components of codeword elements to form $N$-tuples of the MDS code. { Our novel contributions can be summerized as follows:
\begin{itemize}
    \item We propose two novel modulation concepts that we call MDS amplitude and phase modulation (MDS-APM) and MDS in-phase and quadrature modulation (MDS-IQM). We exploit the amplitude and phase levels to form the tuples of a simple MDS code for the former, and in-phase and quadrature components for the latter.\footnote{ We use   the MDS coding mechanism to embed information into the codeword elements in a symbol-wise manner. On top of that, the channel coding techniques could be performed independently in a bit-wise manner to improve the performance of the proposed schemes.}
    \item We depict practical OFDM implementations of the proposed schemes. We further show that the codebooks of these implementations are capable of encompassing those of MM-OFDM-IM scheme as a special case, making the proposed schemes important benchmarks for the OFDM-IM techniques. Thus, they are more general classes of channel encoding which provide high SE and satisfactory error performance.
    \item We investigate the minimum Euclidean distance (MED) and bit error rate (BER) of OFDM with MDS-APM (OFDM-MDS-APM) and OFDM with MDS-IQM (OFDM-MDS-IQM) and obtain an upper-on the BER.
    \item Our numerical, as well as analytical findings, indicate that the proposed schemes are capable of outperforming existing OFDM benchmarks in terms of detection complexity, MED, achievable rate, and error performance.
    \item The studies in the IM literature focus mainly on showing the effectiveness of the IM techniques against the conventional modulation techniques such as QAM and PSK or the OFDM concepts employing these techniques. However, the actual potential of the conventional techniques is vastly ignored in these studies as they do not consider well-known encoding techniques when benchmarking the IM techniques. In this regard, this paper shows that one can achieve better error performance than the IM and related concepts while preserving an encoding and decoding structure as simple as these concepts by using a conventional technique called MDS code.  
\end{itemize}}

The rest of the paper is organized as follows. In Section \ref{sec:section2}, we show the basic idea of MDS coding based modulation and define the concepts, MDS-APM and MDS-IQM. OFDM implementations of MDS-APM and MDS-IQM are described in Sections \ref{sec:section3}.  Performance analysis is undertaken in Section \ref{sec:section4}. We present and compare analytical and simulation results in Section \ref{sec:section5}. Finally, we conclude the paper in Section \ref{sec:section6}.


\section{Proposed Modulation Concepts}\label{sec:section2}

{ It is well-known that one needs to maximize the minimum Hamming distance between the codewords to have a good codebook for fading channels \cite{Tarokh1998,Vucetic2003}. In this context, PM can outperform conventional modulation techniques such as QAM and PSK since the minimum Hamming distance between the symbols of PM is two \cite{Nordio2003}, whereas such a distance is limited to one for conventional modulation symbols. However, the potential of the PM concepts is limited since they are not capable of producing the highest number of codewords whose minimum Hamming distance is two. In this regard, we noticed that one can achieve such a number of codewords by using a simple MDS code in the modulation design with even lower complexity than the PM concepts and outperform them in terms of SE and error performance.\footnote{{ We show the effectiveness of the proposed modulation concepts against the PM concepts in the next sections through the recent OFDM implementations of these concepts.}}}

We use a simple MDS code to form information-bearing symbols. The MDS code constructs the first $N-1$ elements, $I_{\tau},\tau\in \big\{1, \ldots, N-1\big\}$, by using the integers, $1,\ldots,Q$ as symbols where $Q$, i.e., $I_{\tau}\in \big\{1, \ldots, Q\big\}$, and the last symbol, $I_N$ is selected from the same integers by letting the code be those $N$-tuples summing to zero under modulo-$Q$ arithmetic, i.e., $(I_1+I_2+\ldots+I_N) \bmod Q=0 $ \cite{Singleton64}.  Therefore, by using the MDS code, one can form $Q^{N-1}$ $N$-tuples. It is important to note that the minimum Hamming distance between the $N$-tuples of this MDS code is two and $Q^{N-1}$ is the maximum number of $Q$-ary $N$-tuples that satisfy this condition \cite{Singleton64}.  Now, we define two modulation concepts based on this MDS code. 

\subsection{Maximum Distance Separable  Amplitude and Phase Modulation}\label{sec:sec21}

In MDS-APM, codebook consists of $L_1$ codewords and each codeword is an $N$-tuple of complex numbers which can be regarded as a vector $\textbf{x}_{l,v}\coloneqq[x_{l,v}(1), x_{l,v}(2), \ldots, x_{l,v}(N)]$ in an Euclidean space $\mathcal{S}$ of $2N$ dimensions where $x_{l,v}(n)=|x_{l,n}|e^{j\varphi_{v,n}} \in \mathbb{C}$ and $n=1, 2, \ldots, N$. Here, we use the $N$-tuples of the MDS code to determine indices related to the amplitudes and phases of the codeword elements, i.e., $|x_{l,n}|$ and $\varphi_{v,n}$, respectively. Hence, the values of the codebook elements are chosen according to the integers that form $N$-tuples of the MDS code. More explicitly, the integers $1, 2, \ldots, K$ that form an $N$-tuple regarding the MDS code are used to select one of $K$ amplitude levels, $\sqrt{A_{k}}$, $k=1,\ldots, K$, for codeword elements in a way that   $|x_{l,n}|=\sqrt{\frac{K}{E}A_{I_{l,n}}}$ where $E=A_{1}+A_{2}+\ldots+A_{K}$ and the subscript $I_{l,n}\in \big\{1, 2, \ldots, K\big\}$ is the $n$th integer in the $l$th $N$-tuple $\mathcal{I}_l$ related to the MDS code where $\mathcal{I}_l\coloneqq (I_{l,1}, I_{l,2}, \ldots, I_{l,N} )$.\footnote{It is important to note that the term $\frac{K}{E}$ in $|x_{l,n}|$ is for making the average energy of the codeword elements equal to one. One can also check that each codeword element will have the same average energy since each amplitude level will be observed the same number of times in each codeword due to the structure of the MDS code.} Since the MDS code has $K^{N-1}$ $N$-tuples, we have $K^{N-1}$ codewords related to amplitude levels, i.e., $l=1, 2, \ldots, K^{N-1}$. Similarly, the integers $1, 2, \ldots, P$ that form an $N$-tuple regarding the MDS code are used to select one of $P$ phases, $\phi_{p}\in [0,2\pi]$, $p=1,\ldots, P$, for codeword elements in a way that   $\varphi_{v,n}=\phi_{I_{v,n}}$ where
the subscript $I_{v,n}\in \big\{1, 2, \ldots, P\big\}$ is the $n$th integer in the $v$th $N$-tuple $\mathcal{I}_v$ related to the MDS code where $\mathcal{I}_v\coloneqq (I_{v,1}, I_{v,2}, \ldots, I_{v,N} )$. Since the MDS code produces $P^{N-1}$ $N$-tuples, we have $P^{N-1}$ codewords related to these tuples, i.e., $v=1, 2, \ldots, P^{N-1}$. By letting $Q_1\coloneqq KP$, the overall codebook size of MDS-APM is given by $L_1=Q_1^{N-1}$.

\begin{table}[t!]
\centering
\caption{Generating amplitude and phase vectors for MDS-APM when $K=P=2$, and $N=3$.}
\label{tab:table1}
\begin{tabular}{|c|c|c|c|}
\hline
$(N-1)$-Tuple & \begin{tabular}[c]{@{}c@{}}$N$-Tuple\\ MDS Code\end{tabular} & \begin{tabular}[c]{@{}c@{}}Amplitude\\ Vectors\end{tabular} & \begin{tabular}[c]{@{}c@{}}Phase \\ Vectors\end{tabular} \\ \hline
$(1, 1)$    & $\mathcal{I}_1=(1, 1, 2)$                                    & $(\sqrt{2/3}, \sqrt{2/3}, \sqrt{4/3})$                      & $(1, 1,  e^{j\pi})$                            \\ \hline
$(1, 2)$    & $\mathcal{I}_2=(1, 2, 1)$                                    & $(\sqrt{2/3}, \sqrt{4/3}, \sqrt{2/3})$                      & $(1, e^{j\pi}, 1 )$                            \\ \hline
$(2, 1)$    & $\mathcal{I}_3=(2, 1, 1)$                                    & $(\sqrt{4/3}, \sqrt{2/3}, \sqrt{2/3})$                      & $(e^{j\pi}, 1, 1)$                             \\ \hline
$(2, 2)$    & $\mathcal{I}_4=(2, 2, 2)$                                    & $(\sqrt{4/3}, \sqrt{4/3}, \sqrt{4/3})$                      & $(e^{j\pi}, e^{j\pi}, e^{j\pi})$               \\ \hline
\end{tabular}
\end{table}

In Table \ref{tab:table1}, we show am example of how we form the amplitude and phase vectors  for an MDS-APM system when $N=3$, $K=2$, $P=2$, $A_1=1$, $A_2=2$, $\phi_1=0$, and $\phi_2=\pi$. If we have a look at the amplitude vector related to the second $N$-tuple $\mathcal{I}_2=(I_{2,1}, I_{2,2}, I_{2,3})=(1, 2, 1)$, we see that the amplitudes are obtained as $|x_{2,1}|=\sqrt{\frac{K}{E}A_{I_{2,1}}}=\sqrt{\frac{2}{3}}$, $|x_{2,2}|=\sqrt{\frac{K}{E}A_{I_{2,2}}}=\sqrt{\frac{4}{3}}$, and $|x_{2,3}|=\sqrt{\frac{K}{E}A_{I_{2,3}}}=\sqrt{\frac{2}{3}}$ since $A_1=1$, $A_2=2$ and $K/E=2/3$. Moreover, the phase vector related to the same $N$-tuple is given by $\varphi_{2,1}=\phi_1$, $\varphi_{2,2}=\phi_2$, and  $\varphi_{2,3}=\phi_1$ where $\phi_1=0$ and $\phi_2=\pi$. To generate the MDS-APM codewords, we perform element-wise multiplication between the elements of amplitude and phase vectors. Thus, in the case where we use the same $N$-tuple $\mathcal{I}_2$ to generate amplitude and phase vectors, the resulting MDS-APM codeword is given by $\textbf{x}_{2,2}=(\sqrt{2/3},\sqrt{4/3}e^{j\pi}, \sqrt{2/3})$.

In MDS-APM, the mapping of information bits to the amplitude and phase vectors can be performed by using look-up tables. { However, once the transmitted signal vector is received at the receiver, the optimum maximum likelihood (ML) detector performs comparisons between the received signal vectors and the signal vectors available in the look-up tables. Thus, the receiver complexity becomes excessive when the codebook size is large. As will be shown in the next section, an element-wise low complexity detection is possible for the proposed concepts; however, we still need to use look-up tables to map the signal vectors to the information bits and the detection complexity would still be high when the size of the look-up tables is large. To prevent such a high detection complexity, it is crucial to perform the mapping without implementing a look-up table. The structure of the encoding mechanism enables us to perform such a mapping without implementing a look-up table.} The bit-to-symbol mapping for the MDS-APM scheme exhibits a very simple structure since $K$ amplitude levels and $P$ phases can be observed repeatedly on $N-1$ locations and the corresponding $(N-1)$-tuples, which generate the $N$-tuple MDS code, can easily be converted into decimal numbers according to bases $K$ and $P$, respectively. Such decimal numbers related to lexicographically ordered $(N-1)$-tuples are natural numbers that are starting from zero and strictly increasing one by one for such $(N-1)$-tuples just like the lexicographically ordered fixed-length bit sequences. Thus, after mapping each bit sequence to the decimal numbers, we can easily obtain the corresponding $(N-1)$-tuples by using base-$K$ and base-$P$ representations of such decimal numbers with $N-1$ digits. To obtain these $(N-1)$-tuples, we find the integers $\alpha_i$, $i=1, \ldots, N-1$ that satisfy $D=\alpha_1K^0+\alpha_2K^1+\ldots+\alpha_{N-1}K^{N-2}$ where $0 \leq \alpha_i\leq K-1$ and $D$ is the decimal number corresponding to the bit sequence.\footnote{Note that this procedure is equivalent to changing the base of the decimal number. Note also that the same procedure can easily be applied to phase vectors by substituting $P$ into $K$.} Then, the regarding $(N-1)$-tuple is obtained by adding one to $\alpha_i$, i.e., $(\alpha_1+1, \alpha_2+1, \ldots, \alpha_{N-1}+1)$, since the integers used in the MDS code start from one. For the proposed structure, the receiver is also capable of demapping the detected indices to the bit sequences by looking at the first $N-1$ indices denoted by $\hat{I}_{k,1}, \hat{I}_{k,2}, \ldots, \hat{I}_{k,N-1}$. In this case, we subtract one from these indices and obtain the decoded decimal number as $\hat{D}=(\hat{I}_{k,1}-1)K^0+(\hat{I}_{k,2}-1)K^{1}+\ldots+(\hat{I}_{k,N-1}-1)K^{N-2}$. A simple bit-to-index symbol mapping example is shown in Table \ref{tab:table3} for $K=3$ and $N=3$. 

\begin{table}[t!]
\centering
\caption{Bit-to-index mapping for MDS-APM when $K=3$ and $N=3$.}
\label{tab:table3}
\begin{tabular}{|c|c|c|c|c|}
\hline
\begin{tabular}[c]{@{}c@{}}Bit \\ Sequence\end{tabular} & \begin{tabular}[c]{@{}c@{}}Decimal \\ Number\end{tabular} & $(\alpha_1, \alpha_2, \ldots, \alpha_{N-1})$ & $(N-1)$-Tuple & $N$-Tuple \\ \hline
$[0~0~0]$                                               & 0                                                         & (0, 0)                                       & (1, 1)        & (1, 1, 1) \\ \hline
$[0~0~1]$                                               & 1                                                         & (0, 1)                                       & (1, 2)        & (1, 2, 3) \\ \hline
$[0~1~0]$                                               & 2                                                         & (0, 2)                                       & (1, 3)        & (1, 3, 2) \\ \hline
$[0~1~1]$                                               & 3                                                         & (1, 0)                                       & (2, 1)        & (2, 1, 3) \\ \hline
$[1~0~0]$                                               & 4                                                         & (1, 1)                                       & (2, 2)        & (2, 2, 2) \\ \hline
$[1~0~1]$                                               & 5                                                         & (1, 2)                                       & (2, 3)        & (2, 3, 1) \\ \hline
$[1~1~0]$                                               & 6                                                         & (2, 0)                                       & (3, 1)        & (3, 1, 2) \\ \hline
$[1~1~1]$                                               & 7                                                         & (2, 1)                                       & (3, 2)        & (3, 2, 1) \\ \hline
\end{tabular}
\end{table}

\subsection{Maximum Distance Separable In-phase and Quadrature Modulation}

In MDS-IQM, codebook consists of $L_2$ codewords and each codeword is an $N$-tuple of complex numbers which can be regarded as a vector $\textbf{x}_{\rho,\delta}=[x_{\rho,\delta}(1), x_{\rho,\delta}(2), \ldots, x_{\rho,\delta}(N)]$ in an Euclidean space $\mathcal{S}$ of $2N$ dimensions where $x_{\rho,\delta}(n)=x_{\rho,n}^I+jx_{\delta,n}^Q \in \mathbb{C}$ and $n=1, 2, \ldots, N$. Here, the $N$-tuples of the MDS code are used to specify indices related to the reel numbers $x_{\rho,n}^I$ and $x_{\delta,n}^Q$. For the in-phase components of the codeword elements, we use the integers $1, 2, \ldots, R$ that form an $N$-tuple regarding the MDS code to select one of $R$ reel numbers, $x_{r}$, $r=1,\ldots, R$, in a way that  $x_{\rho,n}^I=x_{I_{\rho,n}}$. Here, the subscript $I_{\rho,n}\in \big\{1, 2, \ldots, R\big\}$ is the $n$th integer in the $\rho$th $N$-tuple $\mathcal{I}_{\rho}$ related to the MDS code where $\mathcal{I}_{\rho}\coloneqq (I_{\rho,1}, I_{\rho,2}, \ldots, I_{\rho,N} )$. We also assume $\frac{E_TR}{2}=x_1^2+x_2^2+\ldots+x_R^2$ to make the average energy of the in-phase components equal to $E_T/2$ where $E_T$ is the constraint on average energy of a codeword element. Moreover, we make use of the integers $1, 2, \ldots, T$ to form the quadrature components in the same way as the in-phase components. Thus, $x_{\delta,n}^Q=\Breve{x}_{I_{\delta,n}}$ where $\Breve{x}_t \in \mathbb{R}$ and $t=1, 2, \ldots, T$. The assumption on the average element energy is also valid for the quadrature components, i.e., $\frac{E_TS}{2}=\Breve{x}_1^2+\Breve{x}_2^2+\ldots+\Breve{x}_T^2$.\footnote{Note that the reel numbers that are used to form the in-phase and quadrature components could be the same.} Therefore, the average energy is $E_T$ for the codeword elements of MDS-IQM. Due to the structure of the MDS code, the MDS-IQM codebook has $L_2=Q_2^{N-1}$ codewords where $Q_2=RT$. Thus, $\rho \in \big\{1, 2, \ldots, R^{N-1}\big\}$ and $\delta \in \big\{1, 2, \ldots, T^{N-1}\big\}$.

\begin{table}[t!]
\centering
\caption{Generating in-phase and quadrature components for MDS-IQM when $R=T=2$, and $N=3$.}
\label{tab:table2}
\begin{tabular}{|c|c|c|c|}
\hline
$(N-1)$-Tuple & $N$-Tuple MDS Code        & In-phase Components                       & Quadrature Components                               \\ \hline
$(1, 1)$    & $\mathcal{I}_1=(1, 1, 2)$ & $(\sqrt{2}/2, \sqrt{2}/2, -\sqrt{2}/2)$   & $(1/2, 1/2,  -\sqrt{3}/2)$                \\ \hline
$(1, 2)$    & $\mathcal{I}_2=(1, 2, 1)$ & $(\sqrt{2}/2, -\sqrt{2}/2, \sqrt{2}/2)$   & $(1/2, -\sqrt{3}/2, 1/2 )$                \\ \hline
$(2, 1)$    & $\mathcal{I}_3=(2, 1, 1)$ & $(-\sqrt{2}/2, \sqrt{2}/2, \sqrt{2}/2)$   & $(-\sqrt{3}/2, 1/2, 1/2)$                 \\ \hline
$(2, 2)$    & $\mathcal{I}_4=(2, 2, 2)$ & $(-\sqrt{2}/2, -\sqrt{2}/2, -\sqrt{2}/2)$ & $(-\sqrt{3}/2, -\sqrt{3}/2, -\sqrt{3}/2)$ \\ \hline
\end{tabular}
\end{table}

In Table \ref{tab:table2}, we give an example of how we form the in-phase and quadrature components of an MDS-IQM scheme when $R=T=2$, $N=3$, $x_1=\sqrt{2}/2$, $x_2=-\sqrt{2}/2$, $\Breve{x}_1=1/2$, and $\Breve{x}_2=-\sqrt{3}/2$. When we have a detailed look at the third $N$-tuple of the MDS code, , i.e., $\mathcal{I}_3=(2, 1, 1)$, we see that the elements of in-phase components are $x_{3,1}^I=x_2$, $x_{3,2}^I=x_1$, and $x_{3,3}^I=x_1$, whereas  the elements of quadrature components are $x_{3,1}^Q=\Breve{x}_2$, $x_{3,2}^Q=\Breve{x}_1$, and $x_{3,3}^Q=\Breve{x}_1$. This structure can easily be observed from the regarding row of the table. Moreover, to form the overall codeword related to $\textbf{x}_{2,4}$, for example, we combine the second and forth vectors related to in-phase and quadrature components and obtain it as $\textbf{x}_{2,4}=(\sqrt{2}/2-j\sqrt{3}/2, -\sqrt{2}/2-j\sqrt{3}/2, \sqrt{2}/2-j\sqrt{3}/2)$.  Here, the MDS-IQM concept has $(RT)^{N-1}=16$ codewords. This can also be verified by writing all 16 combinations related to the in-phase and quadrature components.   

\begin{remark}
{
The encoding mechanism of the proposed concepts is substantially different than that of PM in \cite{Slepian65}. The proposed concepts employ an MDS coding mechanism to encode information, whereas the PM concepts use combinatorial tools such as permutations and combinations for the same purpose. Moreover, when there are $N$ different in-phase and quadrature elements for an $N$-element codeword, the MDS-IQM concept can produce $2N^{N-1}$ different codewords. However, in this case, the PM concept can generate $2N!$ different codewords at most. Therefore, the size of the MDS-IQM codebook is always greater than that of the PM codebook when $N>2$. The same result can also be obtained for MDS-APM.} 
\end{remark}

\section{Practical Model For OFDM}\label{sec:section3}

In this section, we show practical applications of MDS-APM and MDS-IQM for OFDM transmissions. The resulting applications are called as OFDM with MDS-APM (OFDM-MDS-APM) and OFDM with MDS-IQM (OFDM-MDS-IQM) for former and latter, respectively. { By these applications, we aim to show an example of how the proposed concepts can be employed in a practical communication scenario. We also aim to conduct fair comparisons with the recent OFDM-IM schemes and show the effectiveness against these schemes in terms of achievable rate and error performance. Besides, a low-complexity detection is shown possible for the OFDM implementations of the proposed concepts, and the complexity of such a detector is compared with those of the OFDM-IM schemes in this section. }

\subsection{Transmitter}

For both systems, $m$ input bits that enter the transmitter are divided into $G$ groups with $f$ information bits where $m=Gf$. Similarly, the total number of subcarriers, $N_T$, is also divided into $G$ groups where each group has $N$ subcarriers, i.e., $N_T=GN$. We use $N$ subcarriers in each group to carry $f$ information bits. The remaining operations are particular for each system and will be explained in the next subsections. { Note also that although we group the subcarriers and information bits in the same way as the OFDM-IM concepts for the sake of low-complexity implementation, the proposed OFDM concepts perform substantially different operations in each group than that of the OFDM-IM concepts. This is mainly because the proposed concepts employ an MDS coding mechanism to encode information into the OFDM signal vectors, whereas the OFDM-IM concepts adapt a combinatorial approach based on the permutations and combinations of the subcarriers. In this way, the proposed OFDM implementations can achieve a higher SE and better error performance than the OFDM-IM and conventional OFDM techniques.}   

Although OFDM-MDS-APM and OFDM-MDS-IQM employ substantially different encoding procedures in each group of bits and subcarriers, they perform the same encoding procedure on their groups. Thus, we will focus on explaining the encoding mechanism for only the $g$th group, $g=1, 2, \ldots, G$, of both systems in the rest of this section. 

\subsubsection{OFDM-MDS-APM}

We further split $f$ information bits into three parts, namely $f_1, f_2$, and $f_3$ with $f=f_1+f_2+f_3$. The first $f_1={\floor{\log_2K^{N-1}}}$ bits are used to determine the amplitudes of the signals carried by $N$ subcarriers, i.e., the energies of the subcarriers, as described in Section \ref{sec:sec21}.  Thus, the integers $1, 2, \ldots, K$ of the MDS code are used to select one of $K$ amplitude levels, $\sqrt{A_k}$, $k=1, \ldots, K$ to map the amplitude of the signal carried by a subcarrier. Thus, $|s_{l,n}^g|=\sqrt{\frac{K}{E}A_{I_{l,n}}}$ where $|s_{l,n}^g|$ is the amplitude of the signal carried by the $n$th subcarrier in the $g$th group and the subscript $I_{l,n}$ is the $n$th integer in the $l$th, $l=1, \ldots, K^{N-1}$, $N$-tuple of the MDS code, i.e., $\mathcal{I}_l$. To form a practical OFDM-MDS-APM scheme, we choose $A_k=k$, thus $|s_{l,n}^g|=\sqrt{\frac{I_{l,n}}{(K+1)/2}}$.\footnote{As will be shown in the Simulation Results Section, such a choice of amplitude levels enables OFDM-MDS-APM to achieve a promising error performance. { It also ensures unit average energy for subcarriers.}}  

Once we map $f_1$ bits to the amplitude of the signals, $f_2={\floor{\log_2P^{N-1}}}$ bits are used to select one of the $P$ disjoint phase sets $\mathcal{M}_p$, $p\in \big\{1, 2, \ldots, P\big\}$, where each set has $M$ different phases, i.e., $|\mathcal{M}_p|=M, \forall p$.\footnote{For two disjoint phase sets $\mathcal{M}_p$ and $\mathcal{M}_{\hat{p}}$ where $p, \hat{p}\in \big\{1, 2, \ldots, P\big\}$ and $p\neq \hat{p}$, $\mathcal{M}_p\cap \mathcal{M}_{\hat{p}}=\emptyset$.} To maximize the angular difference between the constellation points, we obtain the disjoint phase sets $\mathcal{M}_p$ by rotating each constellation with the angle of $2(p-1)\pi/(MP)$, $p=1, 2, \ldots, P$, as in \cite{Yarkin2020, Wen2017}.\footnote{Note that we start forming the disjoint phase sets by using the phases of a conventional $M$-PSK modulation, then we obtain the remaining sets by rotating previous set by $2\pi/(MP)$.}  { In this way, the minimum Euclidean distance between the constellation points is maximized and the best asymptotic BER performance is attained for OFDM-MDS-APM.} To decide the indices of the disjoint sets on the subcarriers, we use the $N$-tuples of MDS code that are formed by the integers $1, 2, \ldots, P$. Then, the remaining $f_3=N\log_2M$ bits are used to determine one of the $M$ phases related to the selected disjoint sets for each subcarrier. These phases form the transmit symbol vectors along with the amplitudes determined by $f_1$ bits. It is important to note that the  OFDM-MDS-APM constellation is similar to that of star-QAM since OFDM-MDS-APM embeds data into both amplitude levels and phases similar to star-QAM whose constellation consists of multiple concentric circles along with the different phase symbols on them. Thus, the amplitude levels in OFDM-MDS-APM can be regarded as circles/rings as in star-QAM. { To maximize the distance between the constellation points on the consecutive circles}, we further rotate each circle by $\pi/(PM)$ degrees compared with the previous circle starting from the one after the innermost circle like the  star-QAM scheme in \cite{Yang2014}.\footnote{{ Although $\pi/(PM)$ is not an optimal rotation angle when $K \neq 2$, we observed from our simulation results that such a rotation angle enables substantially better error performance, at especially low SNR, than the case where there is no rotation.}} The overall data rate per subcarrier for OFDM-MDS-APM is given by 
\begin{align}
    \eta=\frac{\floor{\log_2 K^{N-1}}+\floor{\log_2 P^{N-1}}+N\log_2 M}{N}.
\end{align}


The mapping of $f_1$ bits to the subcarriers' energies and $f_2$ bits to the phase sets can be performed without a look-up table implementation as discussed in Section \ref{sec:sec21}. The symbol vector corresponding to the $g$th group can be written as $\textbf{s}^g=[s_1^g, s_2^g, \ldots, s_N^g]$ where $s_n^g=|s_{l,n}^g|e^{j\varphi_{v,n}^g}$ and $\varphi_{v,n}^g \in \mathcal{M}_{I_{v,n}}$ where the subscripts $v \in \big\{1, 2, \ldots, P^{N-1}\big\}$ and $I_{v,n}\in\big\{1,2,\ldots, P\big\}$. After obtaining the symbol vectors for all groups, an OFDM block creator forms the overall OFDM symbol as $\textbf{s}=[s(1), s(2), \ldots, s(N_T)]^{T}=[\textbf{s}^1, \textbf{s}^2, \ldots, \textbf{s}^G]^T \in \mathbb{C}^{N_T\times 1}$. The remaining operations are the same as those of conventional OFDM.\footnote{We assume that the elements of \textbf{s} are interleaved sufficiently and the
maximum spacing is achieved for the subcarriers. { In this way, the effect of inter-symbol-interference arising from the frequency selectivity of the channel is largely alleviated.}}

\subsubsection{OFDM-MDS-IQM}

In OFDM-MDS-IQM, $f$ information bits are divided into two parts, namely $f_1$ and $f_2$ with $f=f_1+f_2$. We use $f_1$ and $f_2$ bits to determine the in-phase and quadrature components, respectively, of the signals carried by the subcarriers. $f_1$ and $f_2$ bits are further divided into two parts where $f_1=f_{11}+f_{12}$  and $f_2=f_{21}+f_{22}$. Here, we use $f_{11}={\floor{\log_2R^{N-1}}}$ bits to determine the indices related to $R$ disjoint $M$-ary pulse amplitude modulation (PAM) constellations  $\mathcal{P}_r$, $r\in\big\{1, 2, \ldots, R\big\}$.\footnote{To form the disjoint PAM constellations, we use the mode selection strategy in \cite[Sec. IV]{Wen2017}. { As shown in \cite{Wen2017}, such a mode selection strategy is practical and quite efficient in terms of error performance.}} To obtain the indices, we use the $N$-tuples of the MDS code that is formed by the integers $1, 2, \ldots, R$. Assuming $\rho$th, $\rho \in \big\{1, \ldots, R^{N-1}\big\}$, $N$-tuple code is selected for transmission in the $g$th subcarrier/bit group, we relate $I_{\rho, n}$, the $n$th integer in the $\rho$th $N$-tuple, with in-phase component of the signal carried by the $n$th subcarrier and select the in-phase component of the $n$th subcarrier from $\mathcal{P}_{I_{\rho,n}}$. Once we decide the indices of disjoint constellations for all subcarriers, i.e., $I_{\rho,1},I_{\rho,2}\ldots,I_{\rho,N} $, $f_{12}=N\log_2M$ bits are used to select one of $M$-PAM symbols for each subcarrier. The in-phase component of the signal carried by the $n$th subcarrier, $\Re (s_{n}^{g})$, is chosen from the $I_{\rho,n}$th PAM constellation, i.e.  $\Re (s_{n}^{g})=s_{I_{\rho,n}}^g$ where $s_{I_{\rho,n}}^g \in \mathcal{P}_{I_{\rho,n}}$ and $I_{\rho,n}\in\big\{1, 2, \ldots, R\big\} $. Once the mapping of $f_1$ bits to the in-phase components is done, $f_2=f_{21}+f_{22}$ bits are used to determine the quadrature components of the signals in the same way. Hence, we use $f_{21}={\floor{\log_2 T^{N-1}}}$ bits to determine the indices related to $T$ disjoint $M$-PAM constellations, $\mathcal{P}_{t}$, $t\in \big\{1, 2, \ldots, T\big\}$. The $N$-tuple MDS code related to the integers $1, 2, \ldots, T$ is used to determine the indices. Assuming the incoming $f_{21}$ bits corresponds to the $\delta$th, $\delta \in \big\{1, \ldots, T^{N-1}\big\}$, $N$-tuple MDS code, the quadrature component of the $n$th subcarrier is associated with the $I_{\delta,n}$th PAM constellation, $\mathcal{P}_{I_{\delta,n}}$ where $I_{\delta,n}\in \big\{1, \ldots,T\big\}$. Once the indices are determined for all subcarriers, $f_{22}=N\log_2M$ bits are used to select one of $M$-PAM symbols for each subcarrier.\footnote{Note that the PAM constellations that form the in-phase and quadrature components of OFDM-MDS-IQM do not have to be the same in size. However, we assume each constellation consists of $M$ points for simplicity.} Therefore, the quadrature component of the signal carried by the $n$th subcarrier, $\Im (s_{n}^{g})$, is chosen from the $I_{\delta,n}$th PAM constellation, i.e.  $\Im (s_{n}^{g})=s_{I_{\delta,n}}^g$ where $s_{I_{\delta,n}}^g \in \mathcal{P}_{I_{\delta,n}}$ and $I_{\delta,n}\in\big\{1, 2, \ldots, T\big\} $. Thus, the overall data rate per subcarrier for OFDM-MDS-IQM is given by 
\begin{align}
    \eta=\frac{\floor{\log_2 R^{N-1}}+\floor{\log_2 T^{N-1}}+2N\log_2 M}{N}.
\end{align}

    The mapping of $f_{11}$ and $f_{21}$ bits to the disjoint PAM constellations can be performed without implementing a look-up table as discussed in Section \ref{sec:sec21}. The symbol vector corresponding to the $g$th group can be written as $\textbf{s}^g=[s_1^g, s_2^g, \ldots, s_N^g]$ where $s_n^g=s_{I_{\rho,n}}^g+js_{I_{\delta,n}}^g$. After obtaining the symbol vectors for all groups, an OFDM block creator forms the overall OFDM symbol as $\textbf{s}=[s(1), s(2), \ldots, s(N_T)]^{T}=[\textbf{s}^1, \textbf{s}^2, \ldots, \textbf{s}^G]^T \in \mathbb{C}^{N_T\times 1}$. The remaining operations are the same as those of conventional OFDM.

\begin{remark}
When $N$ is odd and we use the integers $1, 2, \ldots, N$ to form the $N$-tuples of the MDS code, the resulting $N$-tuples of the proposed concepts subsume the permutation indices formed by MM-OFDM-IM. This can easily be observed from the fact that $(1+2+\ldots+N)\bmod N=0$ when $N$ is odd. On the other hand, when $N$ is even and we use the integers $1, 2, \ldots, N+1$ to form the $N$-tuples of the MDS code, the resulting $N$-tuples subsume the permutation indices formed by MM-OFDM-IM since $(1+2+\ldots+N) \bmod (N+1)=0$ when $N$ is even. For example, when $N=3$, one can list the $3$-tuples of the MDS code that use the integers $1, 2, 3$ as $(1, 1, 1)$, $(1, 2, 3)$, $(1, 3, 2)$, $(2, 1, 3)$, $(2, 2, 2)$, $(2, 3, 1)$, $(3, 1, 2)$, $(3, 2, 1)$, and $(3, 3, 3)$. Notice that these $3$-tuples subsume all permutations of the set $\big\{1, 2, 3\big\}$.  Thus, the proposed concepts are more general concepts, which
encompass the permutation indices generated by an MM-OFDM-IM encoder. Moreover, the codebooks of the proposed systems are much more flexible than that of MM-OFDM-IM, since the $N$-tuples of the MDS code could be generated by the integers other than $1, 2, \ldots, N, N+1$. 
\end{remark}

\subsection{Receiver}

At the receivers of OFDM-MDS-APM and OFDM-MDS-IQM, the received signal is down-converted. Then,
the cyclic prefix is removed from each received baseband symbol vector before processing with an FFT. After performing a $N_T$ point FFT, the frequency domain received signal vector can be given as
\begin{align}
    \textbf{y}\coloneqq [y(1), y(2), \ldots, y(N_T)]^T=\textbf{S}\textbf{h}+\textbf{n}
\end{align}
where $\textbf{S}=\text{diag}(\textbf{s})$. Here, $\textbf{h}$ and $\textbf{n}$ are $N_T \times 1$ channel and noise vectors, respectively. Their elements are distributed with $\mathcal{CN}(0,1)$ and $\mathcal{CN}(0,N_0)$, respectively, where $N_0$ is the noise variance. 

Since the encoding is performed independently for each group of the OFDM schemes, decoding can also be performed independently at the receivers of them. Thus, we explain the decoding procedures for each subcarrier group of these schemes in the following subsections.

\subsubsection{OFDM-MDS-APM}
 
By using the optimum maximum likelihood (ML) detector, the indices of amplitude levels, ${\mathcal{I}}_l^g$, and disjoint phase sets, ${\mathcal{I}}_v^g$, as well as the phase vector, $\phi_v^g$, related to the transmitted symbols for the $g$th group can be detected as 
\begin{align}\label{eq:eq4}
    (\hat{\mathcal{I}}_l^g, \hat{\mathcal{I}}_v^g, \hat{\mathcal{\phi}}_v^g )= \arg \min_{{\mathcal{I}}_l, {\mathcal{I}}_v, {\mathcal{\phi}}_v } ||\textbf{y}^g-\textbf{S}^g\textbf{h}^g||^2
\end{align}
where $\textbf{y}^g=[y((g-1)N+1),\ldots,y(gN)]^T$, $\textbf{S}^g=\text{diag}(\textbf{s}^g)$, and $\textbf{h}^g=[h((g-1)N+1),\ldots,h(gN)]$.
Moreover, ${\mathcal{\phi}}_v \coloneqq \big\{\varphi_{v,1}, \varphi_{v,2},\ldots, \varphi_{v,N} \big\}  $ is the set of phases drawn from disjoint phase sets with the indices $\mathcal{I}_v$, i.e., $\varphi_{v,n} \in \mathcal{M}_{I_{v,n}}$. The cardinality of this set, $|{\mathcal{\phi}}_v|$, is $M^N$, since we have $M$ phases in each disjoint phase set. 

Since there are $(KP)^{N-1}M^N$ different realizations related to the amplitude levels, disjoint phase sets, and corresponding phases, the number of metric calculations in \eqref{eq:eq4} is  $(KP)^{N-1}M^N$. Thus, the complexity of the optimum ML detector is order of $O((KPM)^{N})$. To overcome this exponential complexity, we employ a low-complexity ML (LC-ML) detector as in \cite{Yarkin2020}. During the LC-ML detection, we order the channel gains of $N$ subcarriers in descending order.\footnote{{ Ordering the channel gains of the subcarriers and decoding the signals carried by the subcarriers with the highest channel gains at first improve the detection performance of the LC-ML detectors greatly since the subcarriers are dependent on each other through an MDS code.}} Except for the subcarrier that has the minimum channel gain, we perform independent ML detection on each subcarrier by choosing one of $K$ amplitude levels, $P$ disjoint phase sets, and $M$ phases that constitutes the closest point (among $KPM$ constellation symbols) to the received signal in two-dimensional Euclidean space. Then, for the remaining subcarrier, we determine the indices of amplitudes and disjoint phase sets in a way that the indices related to the first $N-1$ subcarriers and the remaining subcarrier form an MDS code. Finally, we perform an ML detection on this subcarrier by choosing one of the $M$ phases drawn from the determined phase set. Consequently, the LC-ML detector performs $KPM(N-1)+M$ metric calculations in total, thus the complexity is order of $O(KPMN)$. It is important to note that the complexity of the LC-ML detector scales linearly with $N$, whereas the complexity increases exponentially with $N$ for the optimum ML detector.  

\subsubsection{OFDM-MDS-IQM}

By using the optimum ML detector for OFDM-MDS-IQM, the estimated index sets, $\hat{\mathcal{I}}_{\rho}^g, \hat{\mathcal{I}}_{\delta}^g$, related to the disjoint PAM constellations and sets of PAM symbols, $\hat{{\Gamma}}_{\rho}^g,  \hat{\mathcal{Q}}_{\delta}^g$, drawn from these constellations can be written as  
\begin{align}\label{eq:eq5}
    (\hat{\mathcal{I}}_{\rho}^g, \hat{\mathcal{I}}_{\delta}^g,\hat{{\Gamma}}_{\rho}^g,  \hat{\mathcal{Q}}_{\delta}^g )= \arg \min_{{\mathcal{I}}_{\rho}, {\mathcal{I}}_{\delta}, {{\Gamma}}_{\rho}^g,{\mathcal{Q}}_{\delta}^g } ||\textbf{y}^g-\textbf{S}^g\textbf{h}^g||^2
\end{align}
where ${{\Gamma}}_{\rho}^g \coloneqq \big\{s_{I_{\rho,1}}^g, s_{I_{\rho,2}}^g, \ldots, s_{I_{\rho,N}}^g \big\}$ and ${\mathcal{Q}}_{\delta}^g\coloneqq \big\{s_{I_{\delta,1}}^g, s_{I_{\delta,2}}^g, \ldots, s_{I_{\delta,N}}^g\big\}$ are the sets of PAM symbols whose $n$th elements are drawn from $\mathcal{P}_{I_{\rho,n}}$ and $\mathcal{P}_{I_{\delta,n}}$, respectively. 

Since the OFDM-MDS-IQM codebook consists of $(RT)^{N-1}M^{2N}$ symbol vectors, the number of metric calculations in \eqref{eq:eq5} is $(RT)^{N-1}M^{2N}$. Thus, the complexity of \eqref{eq:eq5} is order of $O((RTM^2)^N)$. Like OFDM-MDS-APM, the optimum ML detection becomes impractical for OFDM-MDS-IQM since the detection complexity is considerably high when the codebook size is large. To overcome this high complexity, we exploit the same detection strategy as the LC-ML detector of OFDM-MDS-APM. Hence, we decide the indices of disjoint PAM sets and related PAM symbols independently for all subcarriers except for the subcarrier corresponding to the minimum channel gain. Then, we determine the indices of disjoint PAM constellations related to the in-phase and quadrature components of the symbol carried by the remaining subcarrier in a way that the estimated indices form an MDS code. Finally, we perform an ML detection for the remaining subcarrier and choose the closest symbol to the received symbol among  $M^2$ symbols drawn from the determined PAM sets. Here, the LC-ML detector performs $(RTM^2)(N-1)+M^2$ metric calculations, thus the complexity is order of $O(RTM^2N)$.

\subsubsection{Detection Complexity}

In this subsection, we compare the detection complexities related to the low-complexity detectors of the proposed schemes with those of the OFDM-IM schemes. We also provide detection complexity of conventional OFDM for benchmarking.

\begin{table}[t!]
\centering
\caption{Detection complexity comparisons for OFDM-MDS-IQM, OFDM-IM, MM-OFDM-IM, and conventional OFDM.}
\label{tab:table4}
\resizebox{\textwidth}{!}{%
\begin{tabular}{|c|c|c|c|c|c|}
\hline
\multirow{2}{*}{}                                                                                & \multicolumn{2}{c|}{\begin{tabular}[c]{@{}c@{}}OFDM-MDS-IQM\\ $(N, R, T, M)$\end{tabular}} & \begin{tabular}[c]{@{}c@{}}OFDM-IM\\ $(N, K_1, M_1)$\end{tabular} & \begin{tabular}[c]{@{}c@{}}MM-OFDM-IM\\ $(N, M_2)$\end{tabular} & \begin{tabular}[c]{@{}c@{}}OFDM\\ $M_3$-PSK/QAM\end{tabular} \\ \cline{2-6} 
                                                                                                 & ML                                                   & LC-ML                                & LLR                                                               & Subcarrier-Wise                                                 & ML                                                         \\ \hline
\begin{tabular}[c]{@{}c@{}}$N=4, R=2, T=2, M=4, K_1=3$,\\ $M_1=102, M_2=23, M_3=46$\end{tabular} & $1.05 \times 10^6$                                   & 52                                   & 102                                                               & 57.5                                                            & 46                                                         \\ \hline
\begin{tabular}[c]{@{}c@{}}$N=8, R=2, T=4, M=4, K_1=7$,\\ $M_1=142, M_2=27, M_3=99$\end{tabular} & $1.13 \times 10^{15}$                                & 114                                  & 142                                                               & 121.5                                                           & 99                                                         \\ \hline
\begin{tabular}[c]{@{}c@{}}$N=16, R=4, T=4, K_1=15$,\\ $M_1=256, M_2=32, M_3=216$\end{tabular}   & $1.33 \times 10^{36}$                                & 241                                  & 256                                                               & 272                                                             & 216                                                        \\ \hline
\end{tabular}%
}
\end{table}

 In Table \ref{tab:table4}, we compare the detection complexities of OFDM-MDS-IQM, OFDM-IM, MM-OFDM-IM, and conventional OFDM  in terms of the number of metric calculations per subcarrier.\footnote{The number of metric calculations of the LC-ML detector will be the same for OFDM-MDS-APM and OFDM-MDS-IQM when the codebooks of these concepts have the same number of index and conventional modulation symbols. Thus, we included only the complexity results related to OFDM-MDS-IQM in Table \ref{tab:table4} for brevity.} Such a number can be given as $(RT)^{N-2}M^{2N}$, $RTM^2-RTM^2/N+M^2/N$, $M_1$, $M_2N/2+M_2/2$, and $M_3$ for the optimum and low-complexity ML detectors of OFDM-MDS-IQM $(N, R, T, M)$, log-likelihood ratio (LLR) detector of OFDM-IM $(N, K_1, M_1)$ \cite{Basar2013}, subcarrier-wise detector of MM-OFDM-IM $(N, M_2)$ \cite{Wen2017}, and the optimum ML detector of conventional OFDM, respectively. Here, OFDM-MDS-IQM $(N, R, T, M)$ signifies the OFDM-MDS-IQM concept with $N$ subcarriers in each subcarrier group as well as $R$ and $T$ disjoint $M$-ary PAM constellations. Moreover, OFDM-IM $(N, K_1, M_1)$ is the OFDM-IM scheme which activates $K_1$ out of $N$ subcarriers in each subcarrier group and carries $M_1$-ary constellation symbols on the activated subcarriers, whereas MM-OFDM-IM $(N, M_2)$ stands for the MM-OFDM-IM concept that uses permutations of $N$ disjoint $M_2$-ary constellations to form the symbols carried by $N$ subcarriers. To send the same number of the information bits for each scheme, thus to make fair comparisons, we adjust $M_1$, $M_2$, and $M_3$, sizes of the constellation in the OFDM-IM, MM-OFDM-IM, and conventional OFDM schemes, respectively. As seen from the table, the LC-ML detector of OFDM-MDS-IQM is capable of exhibiting a lower complexity than the low-complexity detectors of OFDM-IM and MM-OFDM-IM. Moreover, its complexity is comparable to the ML detector of conventional OFDM. It is also worth mentioning that the LC-ML detectors of the proposed schemes do not require the knowledge of the noise variance unlike the LLR detector of OFDM-IM and subcarrier-wise detector of MM-OFDM-IM.  

Also, the number of operations per decoded bit is commonly used as a metric for the computational complexity \cite{Khandekar2001}. Let's denote by $\zeta_p$ the decoding complexity measured in the receiver operations per information bit where the subscript $p$ is used to signify one of the OFDM schemes, i.e., $p\in \big\{\text{OFDM-IM}, \text{MM-OFDM-IM}, \text{OFDM-MDS-APM}\big\}$. Assuming each scheme has the same SE, $\eta$, per subcarrier and number of subcarriers, $N$, in each subcarrier group, the decoding complexity for the OFDM schemes in the question is given by
$\zeta_p=\frac{c_p}{\eta}$ where $c_p$ stands for the complexity in terms of the number of metric calculations per subcarrier. Following the discussion above, the complexities of the OFDM-MDS-APM, MM-OFDM-IM, and OFDM-IM schemes can be written as $c_{\text{OFDM-MDS-APM}}=KPM-KPM/N+M/N$, $c_{\text{MM-OFDM-IM}}=M_2N/2+M_2/2$, and $c_{\text{OFDM-IM}}=M_1$. The SEs of MM-OFDM-IM and OFDM-IM are given by $\eta_{\text{MM-OFDM-IM}}=\frac{\log_2N!+N\log_2M_2}{N}$ and $\eta_{\text{OFDM-IM}}=\frac{\log_2 \binom{N}{K_1}+K_1\log_2M_1}{N}$, respectively. Let's assume that the modulation order employed by each subcarrier is equal for the OFDM-MDS-APM and MM-OFDM-IM schemes, i.e., $M_2=M$. Moreover, to make the SEs of these schemes equal, one should satisfy $\log_2 N!=(N-1)\log_2 (KP)$ , i.e., $KP=(N!)^{1/(N-1)}$. In this way, we ensure that the OFDM-MDS-APM and MM-OFDM-IM techniques have the same SE along with the same number of index and conventional modulation symbols, i.e., $\eta=(1/N)\log_2 N!+\log_2 M$. Let's also choose $K_1=N-1$ for the OFDM-IM scheme as this choice is shown in \cite{Siddiq2017} to achieve a high SE for the OFDM-IM scheme. It is straightforward to show that one should pick the modulation order on the activated subcarriers as $M_1=(M^N(N-1)!)^{1/(N-1)}$ to make the SE of the OFDM-IM scheme equal to $\eta$. Thus, in this case, the decoding complexity per bit for the OFDM-MDS-APM, MM-OFDM-IM, and OFDM-IM schemes can be written respectively as 
\begin{align}
    \zeta_{\text{OFDM-MDS-APM}}&=\frac{(N!)^{1/(N-1)}M(1-1/N)+M/N}{\eta}, \\ 
    \zeta_{\text{MM-OFDM-IM}}&=\frac{MN/2+M/2}{\eta}, \\
    \zeta_{OFDM-IM}&=\frac{(M^N(N-1)!)^{1/(N-1)}}{\eta}.
\end{align}

\begin{remark}\label{prop:prop1}
When $N\to \infty$, the SEs of OFDM-MDS-APM and MM-OFDM-IM can easily be given by $\eta \sim \log_2 KPM$ and $\eta \sim \log_2 M_2 + \log_2 N-\log_2 e$, respectively, where the latter follows from the Stirling's approximation. To achieve the same SE as that of MM-OFDM-IM, one should satisfy $KPM=M_2N/e$. In this case, the decoding complexities can be written as $\zeta_{\text{OFDM-MDS-APM}} \sim \frac{M_2N}{e\log_2 (M_2N/e)}$ and $\zeta_{\text{MM-OFDM-IM}} \sim \frac{M_2N}{2\log_2 (M_2N/e)}$. Assuming the same modulation order for each subcarrier, one can also show the same result for OFDM-MDS-IQM. Thus, for the OFDM-MDS-APM (IQM) and MM-OFDM-IM concepts that achieve the same SE with the same number of subcarriers $N$, the decoding complexity per bit is lower for OFDM-MDS-APM (IQM) when $N\to \infty$.   
\end{remark}

\begin{figure}[t!]
		\centering
		\includegraphics[width=12cm,height=9cm]{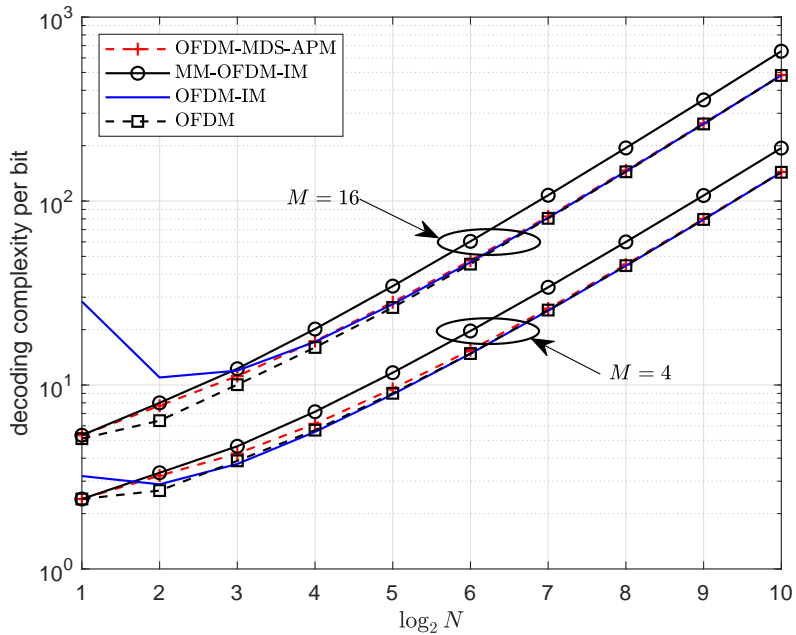}
		\caption{Complexity comparison of OFDM-MDS-APM with MM-OFDM-IM, OFDM-IM, and OFDM in terms of decoding complexity per bit.}
		\label{fig:figtc}
\end{figure}

We provide the complexity values as a function of $\log_2N$ in Fig. \ref{fig:figtc} for $M=4$ and $M=16$. We also plot the decoding complexity of the conventional OFDM in this figure for benchmarking. The figure verifies the result stated in the remark. The proposed OFDM-MDS-APM scheme outperforms the MM-OFDM-IM scheme since it achieves the same SE as that of MM-OFDM-IM by using a fewer number of disjoint constellations. Moreover, its complexity approaches to those of the OFDM-IM and OFDM schemes as $N$ increases.  { It is important to note that although the proposed OFDM schemes employing the LC-ML detectors are slightly outperformed by the conventional OFDM scheme in terms of detection complexity, they exhibit superior error performance compared to conventional OFDM as will be shown in the Simulation Results Section.}

\section{Performance Analysis}\label{sec:section4}

\subsection{Average Bit Error Rate}

In this section, we derive an upper-bound based on the well-known union bound for the average BER of the proposed schemes. As shown in the Simulation Results Section, the derived bound is tight, especially at high SNR. 

Let's define $P(\textbf{S}_i \to \textbf{S}_j)$ as the pairwise error probability (PEP) related to the erroneous detection of $i$th codeword of the codebook, $\textbf{s}_i$, as the $j$th codeword, $\textbf{s}_j$, where $i \ne j $, $i, j \in \big\{1, 2, \ldots, 2^f\big\}$,  $\textbf{S}_i=\text{diag}(\textbf{s}_i)$ and $\textbf{S}_j=\text{diag}(\textbf{s}_j)$. Based on the ML detection rules in \eqref{eq:eq4} and \eqref{eq:eq5}, the PEP conditioned on the channel coefficients, \textbf{h}, can be given as
\begin{align}\label{eq:eq6new}
    P(\textbf{S}_i \to \textbf{S}_j|\textbf{h})=Q\bigg(\sqrt{\frac{\gamma ||(\textbf{S}_i-\textbf{S}_j)\textbf{h}||^2}{2}}\bigg)
\end{align}
where $Q(.)$ is the Q-function and  $\gamma=1/N_0$ is the average received SNR.

By substituting the identity $Q(x)=\frac{1}{12}e^{-x^2/2}+\frac{1}{4}e^{-2x^2/3}$ into \eqref{eq:eq6new} and taking an average over $\textbf{h}$, we obtain the following approximate unconditional PEP expression as in \cite{Basar2013}:
\begin{align}
    P(\textbf{S}_i \to \textbf{S}_j)&=\text{E}_{\textbf{h}}[P(\textbf{S}_i \to \textbf{S}_j|\textbf{h})]
     \\ & \nonumber \approx \frac{1/12}{\text{det}(\textbf{I}_N+\frac{\gamma}{4}\textbf{C}\textbf{A}_{ij})}+\frac{1/4}{\text{det}(\textbf{I}_N+\frac{\gamma}{3}\textbf{C}\textbf{A}_{ij})}
\end{align}
where $\textbf{I}_N$ is the identity matrix of size $N \times N$, $\textbf{C}=\text{E}_{\textbf{h}}[\textbf{h}\textbf{h}^H]$, and $\textbf{A}_{ij}=(\textbf{S}_i-\textbf{S}_j)^H(\textbf{S}_i-\textbf{S}_j)$.

The average BER of the proposed OFDM schemes can be upper-bounded in the same way as in \cite{Basar2013,Wen2017,Yarkin2020} by using the well-known union bound as follows
\begin{align}\label{eq:eq6}
    P_b \le \frac{1}{f2^f}\sum_{i=1}^{2^f}\sum_{j=1}^{2^f} P(\textbf{S}_i \to \textbf{S}_j) D(\textbf{S}_i \to \textbf{S}_j)
\end{align}
where $D(\textbf{S}_i \to \textbf{S}_j)$ is the Hamming distance between bit sequences related to codewords $\textbf{s}_i$ and $\textbf{s}_j$.  

\begin{remark}
The codebooks of OFDM-MDS-APM and OFDM-MDS-IQM consist of $(KP)^{N-1}$ and $(RT)^{N-1}$ codewords related to the indices of constellation points, respectively, and the minimum Hamming distance related to these codewords is two. However, they also include $M^N$ and $M^{2N}$ codewords, respectively, related to the conventional modulation symbols (PSK and PAM symbols for former and latter) that limit the minimum Hamming distance to one. Thus, the diversity order, i.e., the slope of the BER curves, related to the proposed schemes is one. On the other hand, one can comprise only the codewords related to the indices in the codebook, thus achieve a diversity order that is equal to two. In this case, the SEs of the proposed schemes become $\eta=\frac{\floor{\log_2 K^{N-1}}+\floor{\log_2 P^{N-1}}}{N}$ and $\eta=\frac{\floor{\log_2 R^{N-1}}+\floor{\log_2 T^{N-1}}}{N}$ for OFDM-MDS-APM and OFDM-MDS-IQM, respectively. As will be shown in the Simulation Results Section, these special cases of the proposed schemes can achieve a superior error performance at, especially high SNR.  

\end{remark}

\subsection{Minimum Euclidean Distance}

For fading channels, the error performance at low SNR is mainly limited by the minimum Euclidean distances (MEDs) between codeword pairs and the number of codeword pairs that are apart from each other with this distance. However, the MED between the codeword pairs that have minimum Hamming distance in the entire codebook and the number of these pairs limit the error performance at high SNR \cite{Tarokh1998}. In this subsection, we calculate the MEDs of the proposed schemes that are critical measures to assess the error performance. 

\begin{figure}[t!]
		\centering
		\includegraphics[width=9cm,height=9cm]{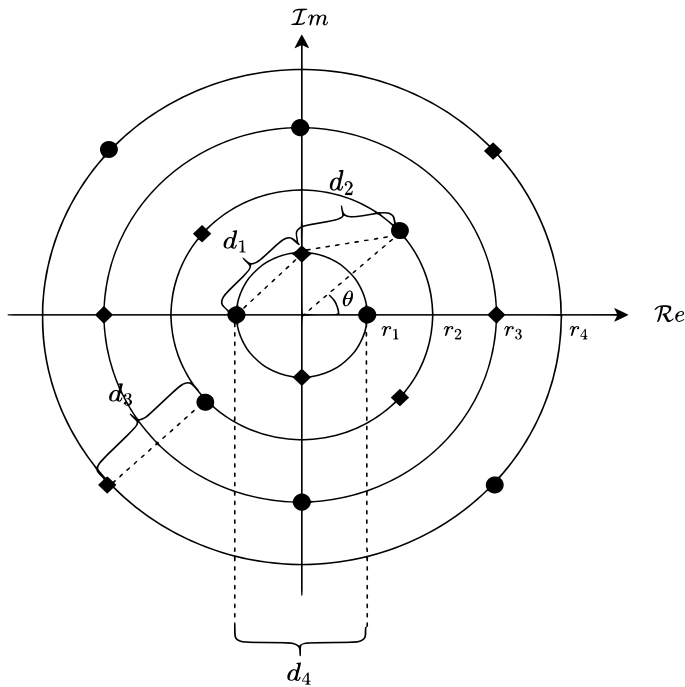}
		\caption{Constellation diagram of OFDM-MDS-APM when $K=4$, $P=2$, and $M=2$.}
		\label{fig:figcons}
\end{figure} 

\subsubsection{OFDM-MDS-APM} As mentioned earlier, OFDM-MDS-APM exhibits similar characteristics to the star-QAM constellation since its codebook consists of multiple concentric circles with phase symbols on them. To make it clearer, Fig. \ref{fig:figcons} illustrates an example of constellation diagram for an OFDM-MDS-APM scheme with $K=4$, $P=2$, and $M=2$. Here, we have $K=4$ amplitude levels ($r_1$, $r_2$, $r_3$, and $r_4$), $P=2$ disjoint phase sets and $M=2$ phase symbols in each set. We denote the elements of one disjoint set by squares, whereas the other is shown by circles. Moreover, the $k$th amplitude level, or in other words, the radius of the $k$th circle is denoted by $r_k$, $k=1,2,3,4$. Note also that we rotate the outer circles by $\theta=\frac{\pi}{PM}=\frac{\pi}{4}$ to increase the distance between the constellation points. As can be seen from the figure, the MED between the constellation points is limited by either $d_1$, the MED between the constellation symbols in the inner-most circle, or $d_2$, the MED between the points in the consecutive circles, or $d_3$, the MED between the points of $\kappa$th and $(\kappa+2)$th circles where $\kappa \in \big\{1, 2, \ldots, K-2\big\}$. However, unlike conventional OFDM, we observe these distances at least two elements of the codeword pairs since the minimum Hamming distance between the index symbols related to the amplitude levels  and disjoint phase sets is two. Thus, the MED of the codebook is limited by the MEDs $\sqrt{2}d_1$,  $\sqrt{2}d_2$, and $\sqrt{2}d_3$. It can also be shown that $d_4$ does not yield the MED as $d_4 \ge \sqrt{2}d_1$.          

In general case where we have $K$ amplitude levels, $P$ disjoint phase sets, and $M$ phases, we can write $r_k=\sqrt{\frac{2k}{K+1}}$ and
\begin{align}
    d_1={\frac{2\sqrt{2}}{\sqrt{K+1}}}\sin ({\pi}/{(PM)}).
\end{align}
In this case, $d_2$ becomes the MED between points in the innermost circles and can be calculated by
\begin{align}
    d_2&=\sqrt{r_1^2+r_{2}^2-2r_1r_{2}\cos({\pi}/{(PM)})}\\ \nonumber & =\sqrt{\frac{6-4\sqrt{2}\cos (\pi/(PM))}{K+1}}.
\end{align}
On the other hand, although rotating circles by $\pi/PM$ maximizes the distance between the points of consecutive circles, it minimizes the distance between the points of $\kappa$th and $(\kappa+2)$th circles where $\kappa \in \big\{1, 2, \ldots, K-2\big\}$ since the constellation points on these circles will have the same phase due to the rotation. It can be shown that the MED between these circles is observed when $\kappa=K-2$, thus such a distance can be given by
\begin{align}
    d_3=\sqrt{\frac{2}{K+1}}\bigg(\sqrt{K}-\sqrt{K-2}\bigg).
\end{align}

By using $d_1$, $d_2$, and $d_3$, the MED of the codebook can be lower bounded by\footnote{Note that $d_{min}$ is a lower bound on the MED of the codebook since the MDS code may not allow us to observe the MEDs $d_1$, $d_2$, and $d_3$ in two elements of the codeword pairs.}
\begin{align}
    d_{min}=\min \big\{\sqrt{2}d_1, \sqrt{2}d_2,\sqrt{2}d_3\big\}. 
\end{align}

On the other hand, the minimum Hamming distance codeword pairs are the unit distance pairs in the disjoint phase sets. It is clear that the MED between those pairs is determined by the pairs in the innermost circle and calculated by
\begin{align}\label{eq:eq15}
    d_4={\frac{2\sqrt{2}}{\sqrt{K+1}}}\sin ({\pi}/{M})
\end{align}
since we have $M$ phases in each disjoint phase set. As mentioned earlier, $d_4$ is the limiting factor for the error performance of OFDM-MDS-APM at high SNR.

\subsubsection{OFDM-MDS-IQM} It can be shown that the MED between the constellation points of OFDM-MDS-IQM is equivalent to the MED between the symbols of a $\xi M$-ary PAM constellations where $\xi = \max \big\{R, T\big\}$ as the $R$-ary and  $T$-ary PAM constellations form the codebook of OFDM-MDS-IQM. Since such a distance is observed in at least two elements of the codeword pairs due to the MDS code, the related MED can be calculated by multiplying the MED of a $\xi M$-ary PAM constellation by $\sqrt{2}$ and obtained as
\begin{align}
    \Breve{d}_{min}=\frac{2\sqrt{3}}{\sqrt{(\xi M)^2-1}}. 
\end{align}
Moreover, the codeword pairs that have the minimum Hamming distance are the $M$-ary symbols in each disjoint PAM constellation. The MED between these symbols can be given by \cite{Wen2017}
\begin{align}\label{eq:eq17}
    \Breve{d}_1=\frac{\sqrt{6}}{\sqrt{M^2-\xi^{-2}}}\sim {\frac{\sqrt{6}}{M}}. 
\end{align}

\subsubsection{Comparison} 
\begin{figure}[t!]
		\centering
		\includegraphics[width=12cm,height=9cm]{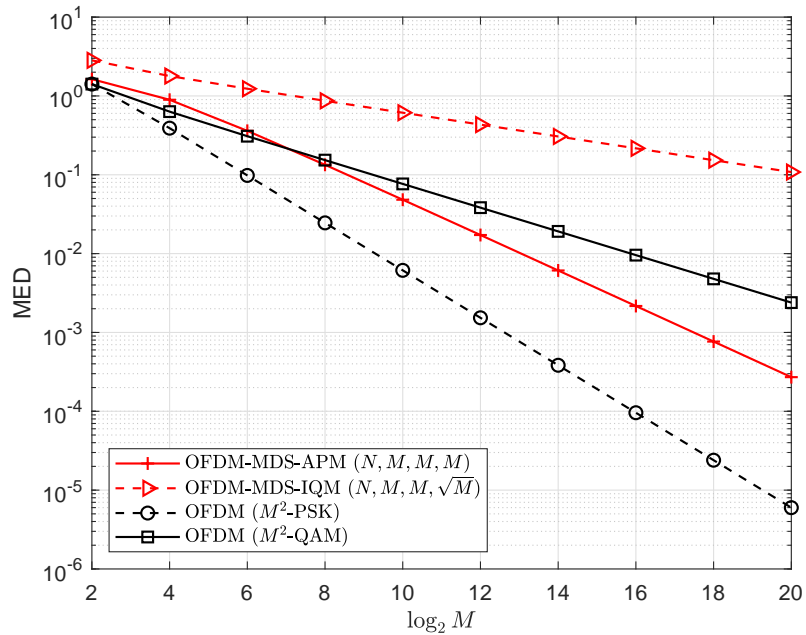}
		\caption{MED comparison of OFDM-MDS-APM $(N, M, M, M)$ and OFDM-MDS-IQM $(N, M, M, \sqrt{M})$  with OFDM ($M^2$-PSK) and OFDM ($M^2$-QAM).}
		\label{fig:figmed}
\end{figure}

To gain more insight into the error performance of the proposed schemes at high SNR, one can compare the MEDs between the minimum Hamming distance codeword pairs for the proposed schemes and conventional OFDM. For comparison purposes, we assume that the proposed OFDM-MDS-APM and OFDM-MDS-IQM concepts achieve an SE that is greater than or equal to the SE of the OFDM benchmarks. To see the impact of both symbols related to the MDS code and conventional modulation symbols in the proposed codebooks, we pick the number of these symbols close to each other. Following these conditions, we consider OFDM-MDS-APM $(N, M, M, M)$ and OFDM-MDS-IQM $(N, M, M, \sqrt{M})$. Then, we compare their MEDs given in \eqref{eq:eq15} and \eqref{eq:eq17} for former and latter respectively, with those of the OFDM schemes that employ $M^2$-PSK and $M^2$-QAM constellations in Fig. \ref{fig:figmed}.\footnote{Here, we calculate the MEDs of OFDM ($M^2$-PSK) and OFDM ($M^2$-QAM) by $2\sin(\frac{\pi}{M^2})$ and $\sqrt{\frac{{6}}{M^2-1}}$, respectively.} In this case, the proposed schemes are capable of achieving a SE that is at least as much as those of the benchmarks when $N\ge2$. As seen from the figure,  OFDM-MDS-IQM $(N, M, M, \sqrt{M})$ achieves a superior MED performance and outperforms the OFDM benchmarks for all values of $M$. Although OFDM-MDS-APM $(N, M, M, M)$ is capable of outperforming OFDM ($M^2$-PSK) and OFDM ($M^2$-QAM), its superiority against OFDM ($M^2$-QAM) disappears when $\log_2(M)>6$. However, the OFDM-MDS-APM scheme can be regarded as an efficient codebook compared to OFDM-IM which can only outperform OFDM in the low-rate range where $\log_2(M)\leq 4$ as shown in \cite{Ishikawa2016}.

\begin{remark}
The proposed concepts achieve a higher MED between the minimum Hamming distance codeword pairs and outperform OFDM in terms of error performance at high SNR. It is also important to note that the proposed schemes can achieve the same SE as those of MM-OFDM-IM in \cite{Wen2017} and OFDM with ordered set partition modulation (OFDM-OFSPM) in \cite{Yarkin2020set} with less number of elements in disjoint constellations. Therefore, the proposed techniques can provide a higher MED than these schemes, leading to a better error performance for OFDM-MDS-APM and OFDM-MDS-IQM at high SNR. We verify these statements in the Simulation Results Section.
\end{remark}

\section{Simulation Results}\label{sec:section5}

In this section, we provide numerical { achievable rate results along with numerical} BER results regarding the optimum ML and LC-ML detectors of the proposed schemes. Moreover, we compare these results with those of the benchmark schemes. We also confirm the theoretical results by computer simulation results in this section.

In figures, ``OFDM-MDS-APM $(N, K, P, M)$'' stands for the proposed OFDM-MDS-APM scheme with $N$ subcarriers in each subcarrier group, $K$ amplitude levels, $P$ phase sets, and $M$ phases in each phase set, whereas ``OFDM-MDS-IQM $(N, R, T, M)$'' is the proposed OFDM-MDS-IQM scheme with $N$ subcarriers in each subcarrier group with $R$ and $T$ disjoint $M$-ary PAM constellations related to the in-phase and quadrature components, respectively. We denote the special cases of the proposed schemes as ``OFDM-MDS-APM $(N, K, P)$'' and ``OFDM-MDS-IQM $(N, R, T)$''. We do not embed information into the conventional modulation symbols for these special cases, thus each disjoint constellation has only one element. Moreover, ``OFDM-IM $(N, K, M$-PSK/QAM)'' stands for the OFDM-IM scheme that activates $K$ subcarriers out of $N$ subcarriers and carries $M$-ary PSK/QAM symbols on the activated subcarriers. ``OFDM-ICM $(N, K, I, M)$'' is the OFDM with index and composition modulation (OFDM-ICM) concept of \cite{Yarkin2021} that activates $K$ subcarriers out of $N$, uses the compositions of an integer $I$ with $K$ parts to embed information into the energies of the activated subcarriers, and carries $M$-PSK symbols on them, whereas, ``OFDM-ICM $(N, K, I)$'' signifies the special case of OFDM-ICM that embeds information into only indices of subcarriers and compositions of an integer $I$ with $K$ parts. ``OFDM-OFSPM ($N$, $M$-PSK/QAM)'' and ``MM-OFDM-IM ($N$, $M$-PSK/QAM)'' stand for the OFDM-OFSPM and MM-OFDM-IM techniques with $N$ subcarriers in each subcarrier group and $N$ disjoint $M$-PSK/QAM constellations to specify the set partitions and permutations for former and latter, respectively. Moreover, ``MM-OFDM-IM-IQ $(N, M)$'' is the MM-OFDM-IM-IQ scheme of \cite{Wen2017} with in-phase and quadrature components of $N$ subcarriers employing $N$ disjoint $M$-PAM constellations, whereas ``$Q$-MM-OFDM-IM ($N$, $Q$, $M$-PSK/QAM)'' represents the $Q$-ary MM-OFDM-IM ($Q$-MM-OFDM-IM) scheme in \cite{Yarkin2020} that employs $Q$ disjoint $M$-PSK/QAM constellations over $N$ subcarriers. Finally, ``OFDM ($M$-PSK/QAM)'' signifies the conventional OFDM with $M$-PSK/QAM. 

In our simulations, we assume all schemes perform transmissions over a Rayleigh fading channel, whose elements are independent and identically distributed. The simulations related to the optimum and suboptimum detectors are executed based on the assumption that channel estimation is perfect. The Rayleigh model assumption is realistic when we consider environments with especially a large number of scatterers \cite{Vucetic2003}. We further assume that the average subcarrier energy is unity for the simulated systems in this section.

\subsection{The Optimum ML Performance}

\begin{figure}[t!]
		\centering
		\includegraphics[width=12cm,height=9cm]{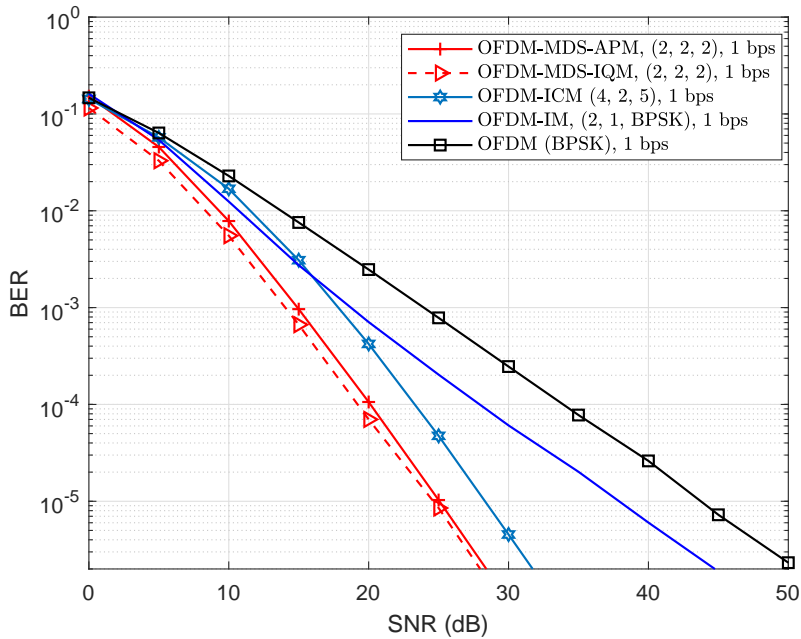}
		\caption{BER comparison of OFDM-MDS-APM (2, 2, 2) and OFDM-MDS-IQM (2, 2, 2) with OFDM-ICM (4, 2, 5), OFDM-IM (2, 1, BPSK), and OFDM (BPSK).}
		\label{fig:fig1}
\end{figure}

In Fig. \ref{fig:fig1}, we compare the BER performance of the proposed OFDM-MDS-APM and OFDM-MDS-IQM schemes with those of the OFDM-ICM, OFDM-IM, and OFDM benchmarks for an SE equal to 1 bit per subcarrier (bps). We perform optimum ML detectors for all schemes of Fig. \ref{fig:fig1}. As seen from the figure, unlike the OFDM and OFDM-IM schemes, the proposed schemes are capable of providing a diversity gain equal to two due to the sole presence of index symbols.\footnote{Here, we call the symbols related to the $N$-tuples of the MDS code as index symbols, since we use the MDS code to embed information into the indices of codeword elements and they have the same minimum Hamming distance as those of the index symbols of the OFDM-IM benchmarks.} They considerably outperform the OFDM and OFDM-IM schemes for all values of SNR illustrated in the figure. They also outperform the OFDM-ICM scheme whose BER curve achieves the same diversity gain as those of the proposed schemes. These results confirm that OFDM-MDS-APM and OFDM-MDS-IQM are promising concepts for cell-edge users in cellular networks as the use of low order modulation is of paramount importance to achieve satisfactory error performance for cell-edge users \cite{Hong2014}. 

\begin{figure}[t!]
		\centering
		\includegraphics[width=12cm,height=9cm]{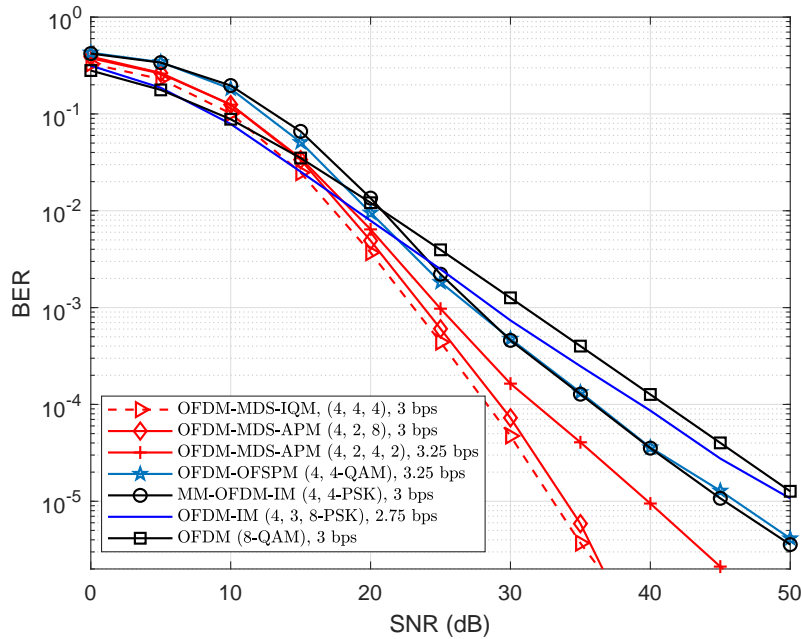}
		\caption{BER comparison of OFDM-MDS-IQM (4, 4, 4), OFDM-MDS-APM (4, 2, 8), and OFDM-MDS-APM (4, 2, 4, 2), with OFDM-OFSPM (4, 4-QAM), MM-OFDM-IM (4, 4-PSK), OFDM-IM (4, 3, 8-PSK), and OFDM (8-QAM).}
		\label{fig:fig2}
\end{figure}

Fig. \ref{fig:fig2} compares the BER performance of the proposed schemes with those of OFDM-OFSPM, MM-OFDM-IM, OFDM-IM, and OFDM. In this figure, the optimum ML detection is considered for all schemes as in Fig. \ref{fig:fig1}. The SEs of the simulated systems range between 2.75 bps and 3.25 bps as shown in the legend of Fig. \ref{fig:fig2}. Here, OFDM-MDS-IQM (4, 4, 4) and OFDM-MDS-APM (4, 2, 8)  achieve a higher diversity order than the benchmarks and outperform them for a wide range of SNR. Although OFDM-MDS-APM (4, 2, 4, 2) is not capable of achieving a higher diversity gain than those of the benchmarks, it considerably outperforms them since it achieves a similar SE to those of the benchmarks by producing a higher number of index symbols and employing a lower modulation order than them.          

\subsection{The Low-complexity Detector Performance}

In this subsection, we compare the BER performance of the proposed schemes with those of the existing OFDM-IM benchmarks by considering the low-complexity subcarrier-wise (LC-SW) detectors of these schemes. Thus, in our simulations, we employ the LC-ML detector for the proposed concepts and the LC-SW detector in \cite{Basar2013} for the OFDM-IM schemes, in \cite{Wen2017} for the MM-OFDM-IM and MM-OFDM-IM-IQ schemes, in \cite{Yarkin2021} for the OFDM-ICM concept, and in \cite{Yarkin2020} for the $Q$-MM-OFDM-IM schemes.  We further provide BER results related to the optimum ML detector of the OFDM scheme for benchmarking. Also, we compare the performance of the LC-ML detector with that of the ML detector.  

\begin{figure}[t!]
		\centering
		\includegraphics[width=12cm,height=9cm]{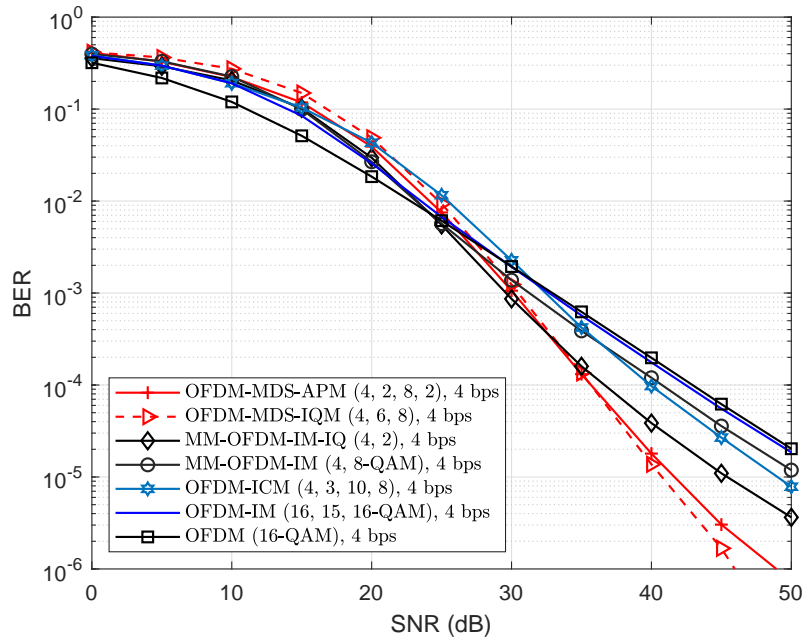}
		\caption{BER comparison of OFDM-MDS-APM (4, 2, 8, 2) and OFDM-MDS-IQM (4, 8, 6) with MM-OFDM-IM-IQ (4, 2), MM-OFDM-IM (4, 8-QAM), OFDM-ICM (4, 3, 10, 8), OFDM-IM (16, 15, 16-QAM), and OFDM (16-QAM).}
		\label{fig:fig3}
\end{figure}

In Fig. \ref{fig:fig3}, we compare the low-complexity detection performance of the proposed systems with those of the MM-OFDM-IM-IQ, MM-OFDM-IM, OFDM-ICM, and OFDM-IM systems when $\eta=4$ bps. As seen from the figure,  although the OFDM-IM benchmarks slightly outperform the proposed concepts at low SNR, our proposals begin to outperform the benchmarks when we increase SNR. Moreover, the SNR loss of the proposed schemes compared to the OFDM scheme is not slight at low SNR like the OFDM-IM benchmarks. However, OFDM-MDS-APM (4, 2, 8, 2) and OFDM-MDS-IQM (4, 8, 6) considerably outperform OFDM (16-QAM) at mainly high SNR. The behavior at the low and high SNR can be explained by the MED between the codeword pairs and the MED between the minimum Hamming distance codeword pairs, respectively. The presence of index symbols for the proposed schemes decreases the overall MED between the codeword pairs, leading to an SNR loss at low SNR. However, it increases the MED between the minimum Hamming distance codeword pairs as the codebook structure enables us to perform a lower order modulation on the subcarriers. The proportion of index symbols to the conventional modulation symbols is higher in the proposed schemes than those of the OFDM-IM benchmarks. That explains the BER improvement compared to these benchmarks at high SNR. As a result, the proposed codebooks yield a trade-off between the performance at low and high SNR. 

\begin{figure}[t!]
		\centering
		\includegraphics[width=12cm,height=9cm]{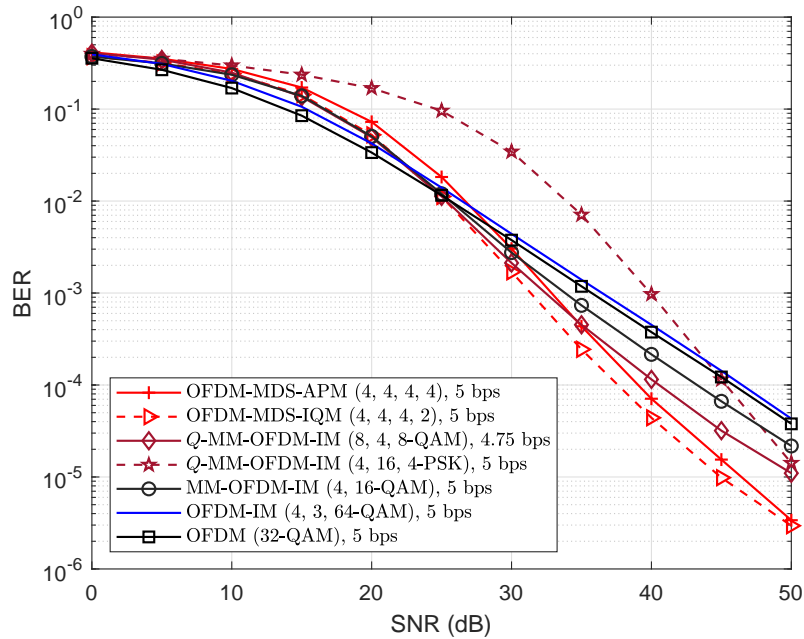}
		\caption{BER comparison of OFDM-MDS-APM (4, 4, 4, 4) and OFDM-MDS-IQM (4, 4, 4, 2) with $Q$-MM-OFDM-IM (8, 4, 8-QAM), $Q$-MM-OFDM-IM (4, 16, 4-PSK), MM-OFDM-IM (4, 16-QAM), OFDM-IM (4, 3, 64-QAM), and OFDM (32-QAM).}
		\label{fig:fig4}
\end{figure}

In Fig. \ref{fig:fig4}, we compare the low-complexity detection performance of the proposed schemes with those of the $Q$-MM-OFDM-IM, MM-OFDM-IM, and OFDM-IM schemes where all concepts provide an SE of 5 bps except for $Q$-MM-OFDM-IM (8, 4, 8-QAM) which achieves an SE of 4.75 bps. Like Fig. \ref{fig:fig3}, the performance of the proposed schemes is superior against those of the OFDM-IM and OFDM benchmarks at especially high SNR. Different than Fig. \ref{fig:fig3}, we provide simulation results related to the $Q$-MM-OFDM-IM systems that are shown in \cite{Yarkin2020} to have superior error performance against the existing OFDM-IM benchmarks. In this regard, the use of amplitude levels in addition to the phase sets enables OFDM-MDS-APM (4, 4, 4, 4) to achieve a considerably better error performance than $Q$-MM-OFDM-IM (4, 16, 4-PSK) that incorporates only phase sets for information encoding. Moreover, both OFDM-MDS-IQM (4, 4, 4, 4) and OFDM-MDS-IQM (4, 4, 4, 2) can outperform $Q$-MM-OFDM-IM (8, 4, 8-QAM) although they achieve a higher SE than $Q$-MM-OFDM-IM (8, 4, 8-QAM). 

\begin{figure}[t!]
		\centering
		\includegraphics[width=12cm,height=9cm]{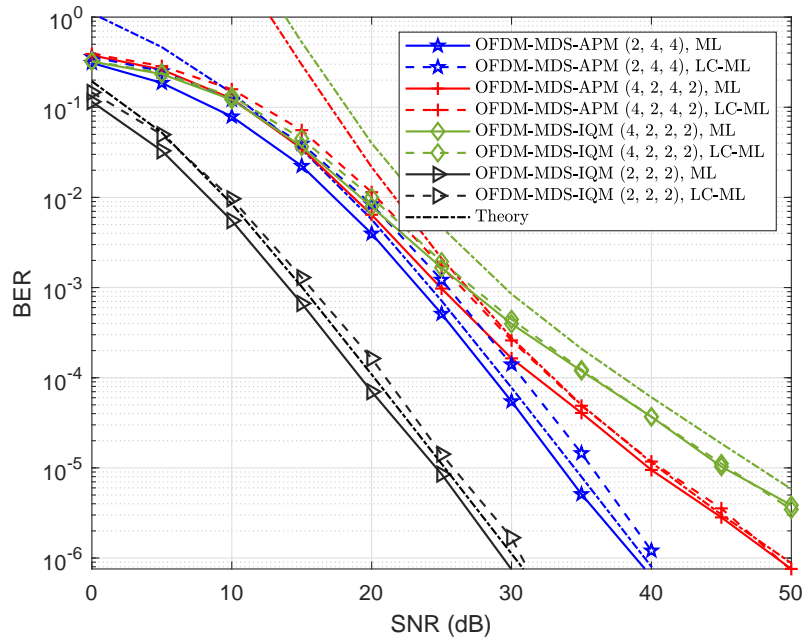}
		\caption{BER comparison of the LC-ML detector with the optimum ML detector for OFDM-MDS-APM (2, 4, 4), OFDM-MDS-APM (4, 2, 4, 2), OFDM-MDS-IQM (4, 2, 2, 2), and OFDM-MDS-IQM (2, 2, 2).}
		\label{fig:fig5}
\end{figure}

In Fig. \ref{fig:fig5}, we compare the BER performance of the LC-ML detector with that of the optimum ML detector for the proposed schemes with different values of SE. The figure demonstrates that the performance of the LC-ML detector is considerably close to that of the ML detector at especially high and low SNR. Although the LC-ML detection results in a slight SNR loss at mid SNR, it is capable of considerably decreasing the complexity compared to the optimum ML detection. Also, the curves related to the label ``Theory'' illustrate theoretical upper-bound results for the average BER. As observed from the figure, upper-bound curves are consistent with computer simulations, especially at high SNR.  

{\subsection{ Achievable Rate}}

{
To gain a practical insight into the achievable rate as well as coded performance of the proposed schemes, one can use the same approach in \cite{Ishikawa2016} and compute the achievable rate numerically by assuming equally likely codewords and evaluating
\begin{align}
    R=\frac{1}{N}\Bigg(f-\frac{1}{2^f}\sum_{i=1}^{2^f}\text{E}_{\textbf{h},\textbf{n}}\bigg[\log_2 \sum_{j=1}^{2^f}e^{\lambda(i, j)}\bigg]\Bigg)
\end{align}
where $\lambda(i, j)=\frac{-||\text{diag}(\textbf{h})(\textbf{s}_i-\textbf{s}_j)+\textbf{n}||^2+||\textbf{n}||^2}{N_0}$. In this subsection, we use this approach to compare the achievable rate of the proposed schemes with that of the benchmarks. }

\begin{figure*}[t!]
    \centering
    \begin{subfigure}[t]{0.5\textwidth}
        \centering
        \includegraphics[width=8.5cm]{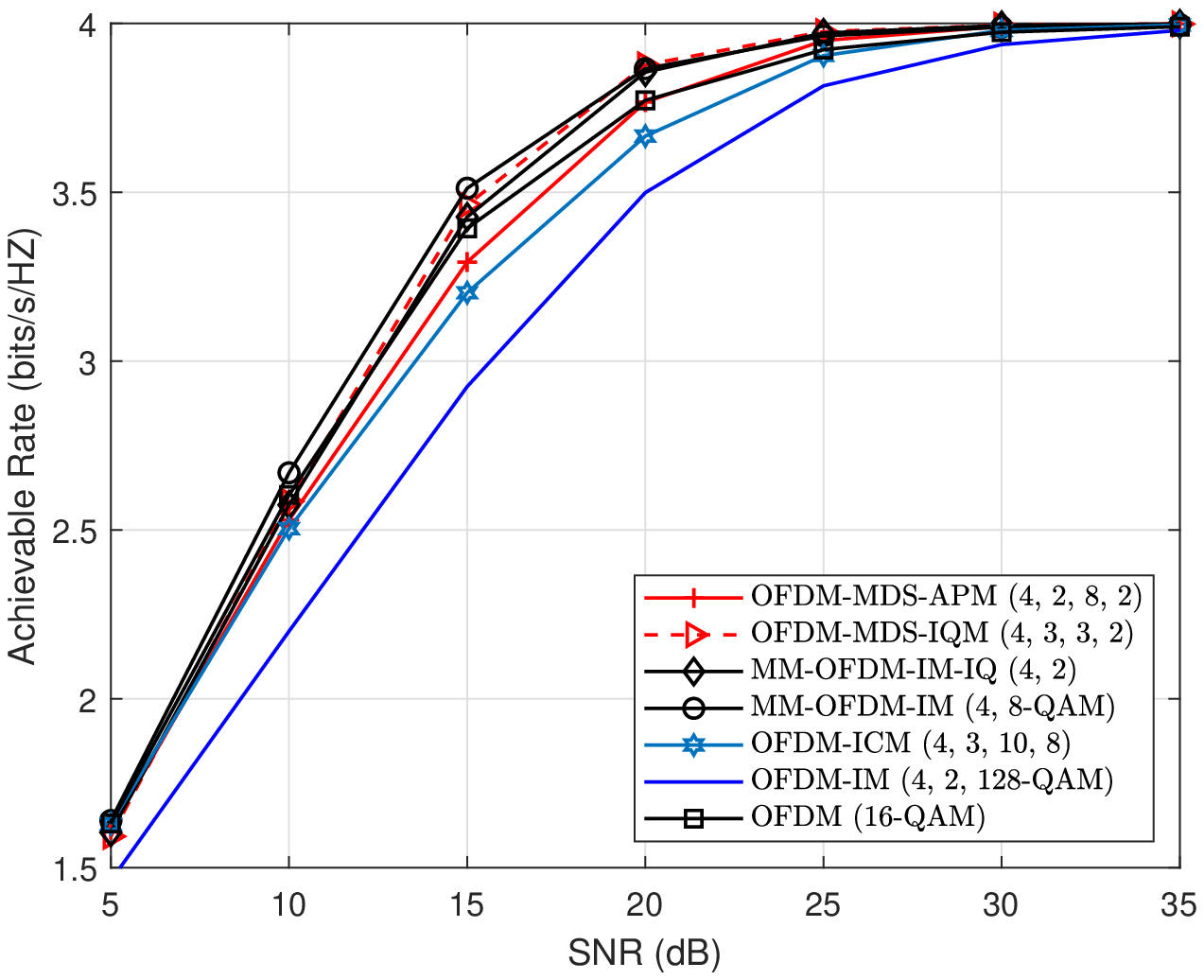}
        \caption{$\eta=4$ bps}
        \label{fig:fig6a}
    \end{subfigure}%
~
    \begin{subfigure}[t]{0.5\textwidth}
        \centering
        \includegraphics[width=8.5cm]{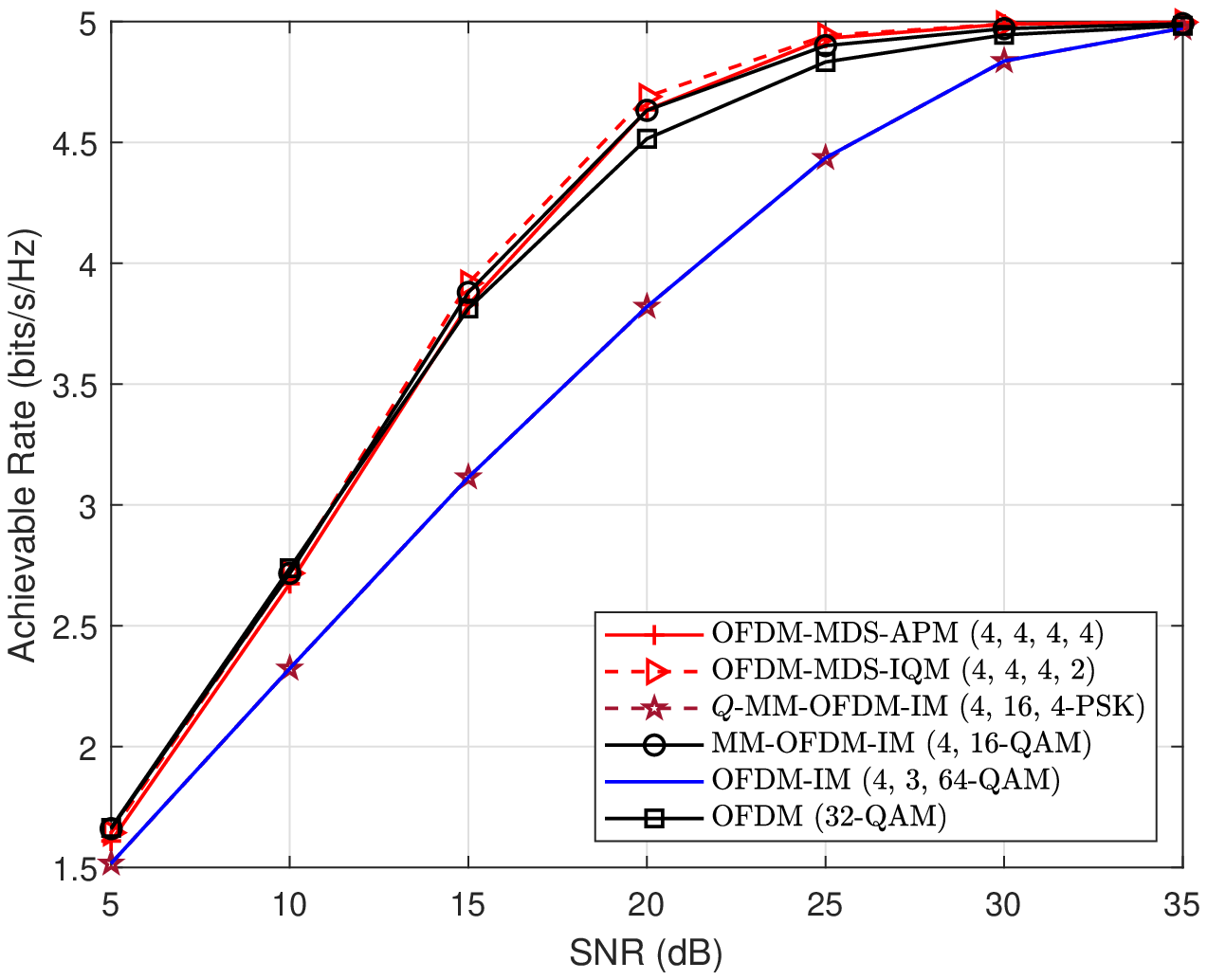}
        \caption{$\eta=5$ bps}
        \label{fig:fig6b}
    \end{subfigure}
    \caption{{ Achievable rate comparison of the proposed and the existing OFDM concepts.}}
    \label{fig:fig6}
\end{figure*}

{ In Figs. \ref{fig:fig6a} and \ref{fig:fig6b}, we compare the achievable rate performance of the proposed OFDM techniques with that of the OFDM benchmarks when the SE is 4 and 5 bps, respectively. As seen from the figures, the proposed OFDM-MDS-IQM schemes outperform all the benchmarks, except for the MM-OFDM-IM scheme when the SE is 4 bps, in terms of achievable rate for a wide range of SNR. Moreover, although the OFDM-MDS-APM scheme is outperformed by the MM-OFDM-IM and OFDM schemes for the SE of 4 bps, it performs closely to the MM-OFDM-IM scheme and  exhibits superior performance  compared to the $Q$-MM-OFDM-IM, OFDM-IM, and OFDM schemes when the SE is 5 bps. It is also important to note that one can improve the performance of the OFDM-MDS-APM technique by carefully choosing the ring ratios and rotation angles related to the constellation of this technique. However, this is beyond the scope of this paper, and it could be considered as future work.}

\section{Conclusion}\label{sec:section6}
In this paper, we proposed two novel modulation concepts that use constellation points to form the codewords of a simple MDS code. { We showed that the proposed concepts are substantially different than the PM concepts, and they are capable of producing more codewords as well as achieving a higher SE. To show an example of practical application and conduct fair comparisons with the recent OFDM-IM techniques, we depicted the OFDM implementations of our modulation designs.} For these implementations, efficient low-complexity decoding and very simple bits-to-symbols/symbols-to-bits mappings were shown possible. { We further compared the detection complexity and MED of the proposed techniques with those of the OFDM benchmarks and demonstrated the effectiveness of our techniques against the benchmarks. Our findings revealed that the proposed concepts are practical, flexible, and low-complexity modulation concepts that are capable of considerably outperforming the existing OFDM benchmarks in terms of error and achievable rate performance.}

As future work, the channel coding techniques could be employed in the proposed concepts to further improve the error performance.  

\bibliographystyle{IEEEtran}
\bibliography{paper}
\end{document}